\documentclass[superscriptaddress, twocolumn, prb, bibnotes, aps, longbibliography, floatfix]{revtex4-2}
\usepackage{comment, siunitx}
\usepackage[utf8]{inputenc}
\usepackage{graphicx}
\usepackage{afterpage}
\usepackage{amsmath, amssymb, bbm, bm}
\usepackage{xcolor}
\usepackage{float,placeins}

\begin{document}

\title{Twisted fibre: a photonic topological insulator}

\author{Nathan Roberts}
\affiliation{Department of Physics, University of Bath, Claverton Down, Bath BA2 7AY, UK}
\affiliation{Centre for Photonics and Photonic Materials, University of Bath, Bath BA2 7AY, UK}

\author{Brook Salter}
\affiliation{Department of Physics, University of Bath, Claverton Down, Bath BA2 7AY, UK}
\affiliation{Centre for Photonics and Photonic Materials, University of Bath, Bath BA2 7AY, UK}

\author{Jack Binysh}
\affiliation{Department of Physics, University of Bath, Claverton Down, Bath BA2 7AY, UK}
\affiliation{Institute of Physics, Universiteit van Amsterdam,
Amsterdam, the Netherlands}

\author{Peter J.~Mosley}
\email{p.mosley@bath.ac.uk}
\affiliation{Department of Physics, University of Bath, Claverton Down, Bath BA2 7AY, UK}
\affiliation{Centre for Photonics and Photonic Materials, University of Bath, Bath BA2 7AY, UK}

\author{Anton Souslov}
\email{as3546@cam.ac.uk}
\affiliation{Department of Physics, University of Bath, Claverton Down, Bath BA2 7AY, UK}
\affiliation{TCM Group, Cavendish Laboratory, JJ Thomson Avenue,
Cambridge, CB3 0HE, UK}

\begin{abstract}
The breaking and enforcing of symmetries is a crucial ingredient in designing topologically robust
materials. While magnetic fields can break time-reversal symmetry to create Chern insulators
in electronic and microwave systems, at optical frequencies natural materials cannot respond to
magnetic fields, which presents a challenge for the scalable exploitation of topologically enhanced devices. Here, we leverage the natural
geometry of fibre to build a scalable photonic Chern insulator by twisting the fibre during
fabrication. The twist inside optical fibre breaks an effective time-reversal symmetry and induces a pseudo-magnetic field, which we observe via photonic Landau levels.
Unavoidably, this twist introduces a competing topology-destroying effect through a parabolic profile in the effective refractive index.
Using simulations to guide experimental materials design, we
discover the Goldilocks regime where the real-space Chern invariant survives, guaranteeing topological protection against fabrication-induced disorder of any symmetry class.
\end{abstract}
\maketitle

Topological band structure engineering enables the creation of robust propagating states in a variety of media~\cite{Hasan2010}. 
In electronic~\cite{vKlitzing1986,macdonald1984} and microwave systems~\cite{Wang2008,haldane2008}, magnetic fields are used to break time-reversal symmetry and create Chern insulators. 
These topological materials act as insulators in the bulk, but display quantised modes at the boundary, which are robust against disorder and localisation~\cite{Hasan2010,TKNNinvariant}. However, materials do not respond to magnetic fields at optical frequencies, creating a challenge for photonic Chern insulators.

 Topology has previously been exploited without breaking time-reversal symmetry through quantum spin Hall~\cite{kane2005, Khanikaev2013, Hafezi2011,umucalilar2011,Hafezi2013} and valley Hall~\cite{Shalaev2019,Ma_Shvets_2016,Dong2016,Chen2017} effects. Unlike Chern insulators, these systems are robust only against select types of disorder~\cite{Rosiek2023,Rechtsman_2023}, but still offer some protection, for example to entangled quantum states~\cite{blanco-redondo2018,Blanco-Redondo2020}. 
An alternative approach to exploit non-trivial topology is to use time-varying metamaterials with a Floquet state~\cite{kitagawa2010,Lindner_2011, Rechtsman2013,Jörg_2017,Yang_2020}. However, the necessary fabrication processes, for example direct-written photonic waveguides, leave these platforms limited in size, flexibility, and scalability. Although optical fibre has been hypothesised to exploit topology on the largest scales~\cite{Huang:23,Lu2018topological,Lin2020,Pilozzi2020, Gong2021,Makwana2020hybrid}, previously fabricated topological fibre~\cite{Roberts2022,Roberts2024} did not break time-reversal symmetry, and was only topologically robust against sub-classes of disorder.

\begin{figure*}[tbp]
\centering
\includegraphics[width=\linewidth]{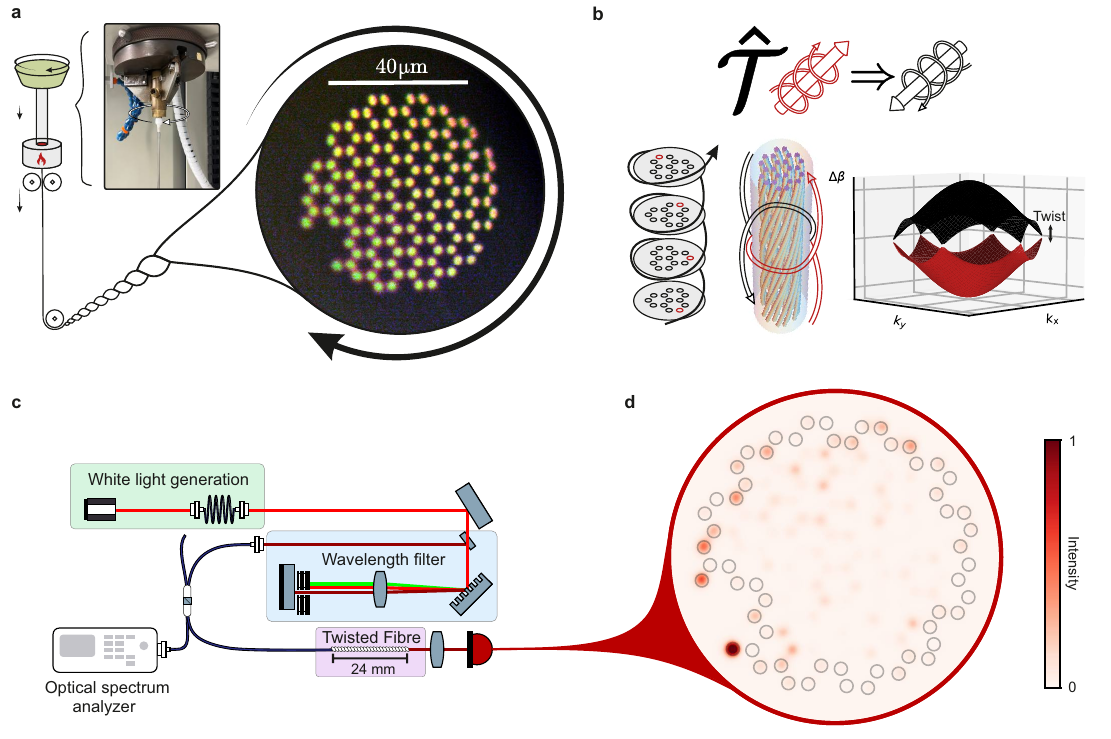}
\caption{Twisted optical fibre overview. \textbf{a}, Diagrammatic explanation of the drawing process. We rotate the preform while feeding it into the furnace to induce a twisted structure. The fibre cross-section features a notch to demonstrate that a topological mode can be guided around an arbitrarily shaped edge. The inset photograph shows a preform being twisted and the micrograph shows the fabricated fibre cross-section with light coupled into the Ge-doped cores. \textbf{b}, Fibre twist schematic showing the breaking of time-reversal symmetry $\hat{\mathcal{T}}$ (equivalently, $z$-propagation symmetry) due to the applied twist. The band structure shows Dirac points in the untwisted system which are gapped when twist is introduced. \textbf{c}, Experimental setup for observing the intensity profile of an excited edge. We use a supercontinuum-generating fibre as a source of white light, which we filter, and butt-couple light at a desired wavelength into our twisted fibre. \textbf{d}, Light has been injected into a single core on the edge of the fibre and propagated over a length of \SI{24}{mm}. The injection core has been overexposed in this image to highlight the intensity distribution in the other excited cores and to show where light is initially injected (c.f.~Fig.~\ref{fig:2}). The edge cores are outlined for reference.}
\label{fig:1}
\end{figure*}

By including many Germanium-doped cores inside the cross-section of a single optical fibre, we build a honeycomb lattice that supports collective supermodes. We engineer the band structure to contain Dirac-point band crossings, and twist the fibre to open a band gap characterised by a Chern topological invariant. We augment the usual stack-and-draw fabrication process by spinning the fibre to freeze-in a constant twist (Fig.~\ref{fig:1}\textbf{a}), which breaks time-reversal symmetry (Fig.~\ref{fig:1}\textbf{b}) and induces experimentally observable robust edge-localised modes (see Fig.~\ref{fig:1}\textbf{c}-\textbf{d}). 
Heuristically, the Chern invariant changes from a non-trivial value inside the lattice of cores to a trivial value in the cladding. This change necessitates the local closing of a topological band gap, leading to modes pinned at the edge. This topological argument demonstrates that the edge modes can be engineered to be of arbitrary shape while remaining robust against disorder without any restrictions on geometry or symmetry.
Neither pseudo-magnetic fields nor this topological class of TRS-breaking have been previously realised in fibre.

To see the unique way in which topology characterises twisted fibre, we begin with the Helmholtz equation for light propagation in a medium with refractive index $n(x,y,z)$:
\begin{equation}
     \nabla^2\mathbf{E} + k^2 n^2\mathbf{E} = 0,\label{Eq:VWE}
\end{equation}
where $k = 2\pi/\lambda$ is the free-space wavevector and $\nabla$ is the three-dimensional gradient. We now decompose the electric field $\mathbf{E} = (\psi_{x},\psi_{y},\psi_{z})e^{i(\beta z - \omega t)}$ into the rapidly oscillating complex phase $e^{i(\beta z - \omega t)}$ with propagation constant $\beta$, and the slowly varying envelope $\psi_i(x,y,z)$. 
Twisted fibre has an inherently $z$-dependent index $n(x,y,z)$, but the transformation into the co-twisting frame,
\begin{align}
    x \rightarrow x \cos(\alpha z) - y \sin(\alpha z)
    \\
    y \rightarrow x \sin(\alpha z) + y \cos(\alpha z),
\end{align}
 creates a  $z$-independent profile $n(x,y)$, where $\alpha$ is the twist rate in radians per meter.
However, this coordinate change introduces several effects inherent to optics in twisted frames. Optical activity mixes the two linear polarisations, resulting in circularly polarised eigenstates $\psi^{\pm}$ with propagation constant splitting $\pm \tau$. As discussed in the Supplementary Information (SI), the torsion $\tau \approx \alpha$ is constant for our small values of twist $\alpha$ and cross-section radius $R$~\cite{ross,Russell_2017,geometro}, so that optical activity can be absorbed into the eigenvalue for each polarisation state.

For each component $\psi = \psi^{\pm}$, we write the propagation equation using the paraxial approximation in the co-twisting frame,
\begin{align}
    i\partial_{z}\psi &=- \frac{1}{2\beta}(\nabla_\perp + i\mathbf{A})^2\psi -\frac{\alpha^2\beta r^2}{2}\psi - \Delta n(x,y) k \psi,
    \label{eq:scalarwaveeq}
\end{align}
where $\nabla_\perp$ is the two-dimensional gradient 
and $\Delta n(x,y)$ is the change in refractive index relative to the silica glass cladding. 
This equation is analogous to the time-dependent Schr\"{o}dinger equation for a charged particle in a constant magnetic field with vector potential $\mathbf{A} = \alpha\beta(y,-x)$ and an external harmonic potential $\alpha^2\beta r^2/2$, where $r \equiv \sqrt{x^2+y^2}$ is the radial coordinate. In this analogy,  dynamics in time correspond to propagation along the $z$-direction, and the eigenstate energy corresponds to the relative propagation constant $\Delta\beta$. The twist then breaks time-reversal symmetry $\hat{\mathcal{T}}$, so traveling backwards along the $z$-direction is equivalent to reversing the sign of $\alpha$: $\hat{\mathcal{T}} (z,\alpha,i) \equiv (-z,\alpha,-i) = (z, -\alpha, i)$, see Fig.~\ref{fig:1}\textbf{b}. 
 In the SI, we use Maxwell's equations in helicoidal coordinates and the paraxial approximation to derive Eq.~(\ref{eq:scalarwaveeq}) and discuss its symmetry properties.

\begin{figure*}[tp]
    \centering
    \includegraphics[width=0.75\linewidth]{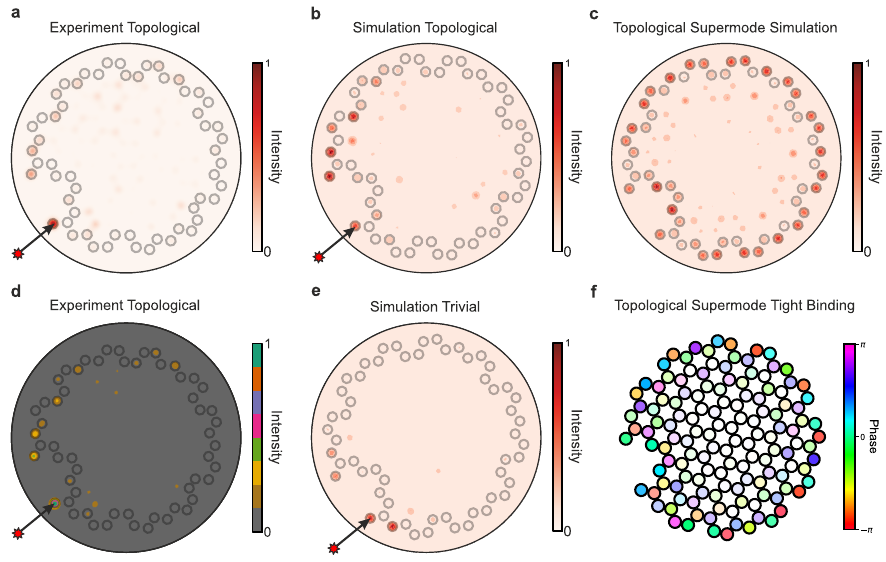}
    \caption{Edge localisation of intensity in topological fibre. \textbf{a}, Experimental image of intensity output after light is injected into the core (marked by the arrow) and propagated through \SI[separate-uncertainty = true]{24.0\pm0.5}{mm} of fibre. (Data from Fig.~\ref{fig:1}\textbf{d} without overexposure.) \textbf{b}, Finite-element simulation of intensity output after light is injected into the marked core and propagated through \SI{23.6}{mm} of fibre. Although the exact profile is highly dependent on the details of the fabrication, the simulations confirm the experimentally observed edge localisation. \textbf{c}, Topological eigenmode at the edge of our twisted fibre obtained from a finite-element simulation.  \textbf{d}, Discretised colourmap of \textbf{a}, showing that the intensity distribution is localised at the edge.  \textbf{e}, Finite-element simulation of intensity output after light is injected into the marked core and propagated through \SI{23.6}{mm} of untwisted, topologically trivial fibre. Instead of propagating around the edge, light remains localised around the initial injection site. \textbf{f}, Topological edge mode calculated from tight-binding numerics, where we plot both the phase (using hue) and the intensity (using saturation). The tight-binding numerics show that the intensity of the edge mode remains localised to the perimeter,  and that the phase varies azimuthally, as expected for Orbital Angular Momentum (OAM) modes.}
        \label{fig:2}
\end{figure*}

In the honeycomb lattice of our twisted fibre cross-section, we describe the band structure for supermodes of $M$ coupled cores using coupled mode theory (equivalent to the tight-binding model). Crucially, the vector potential $\mathbf{A}$ introduces a complex Peierls phase to the (real) coupling strength $C$:
\begin{equation}\label{eq:peierlseqn}\Delta \beta_{m} u_{m} = \sum_{j\neq m} e^{i\mathbf{A}\cdot \mathbf{r}_{mj}} C_{mj} u_{j} + D_{m} u_{m},
\end{equation}
where $u_{m}$ is the contribution from the $m$-th core, $\mathbf{r}_{mj}$ is the position vector between the $m$-th and $j$-th cores, $D_{m}$ is the change in propagation constant for a single core due to the twist, $\Delta \beta_{m}$ is the change in propagation constant for a supermode, and $C_{mj}$ is the coupling strength between neighbouring cores. This coupled mode theory approach is described in the SI and serves as an analytical model for understanding light propagation in our fibre.

Equations~(\ref{eq:scalarwaveeq}--\ref{eq:peierlseqn}) demonstrate the challenges of characterising topology in Chern insulators with broken periodicity, which occur not only in our twisted fibre but also in rotating mechanical lattices~\cite{Cooper2008,Bloch2008,Baksmaty2005,Bhat2007,Williams2010} and cold atomic gases~\cite{Wang_2015,Kariyado_2015,Mitchell2021}.
Although the presence of a periodic honeycomb lattice suggests that solutions for modes $\psi$ could be found in reciprocal space, the simultaneous presence of the effective vector potential $\mathbf{A}$ and the scalar potential $\alpha^2\beta r^2/2$ destroys translational symmetry. Nevertheless, we experimentally and numerically observe topological edge states and use Kitaev summation~\cite{KITAEV20062} to calculate a local topological marker for our system.

\begin{figure*}[tp]
    \centering
    \includegraphics[width=\linewidth]{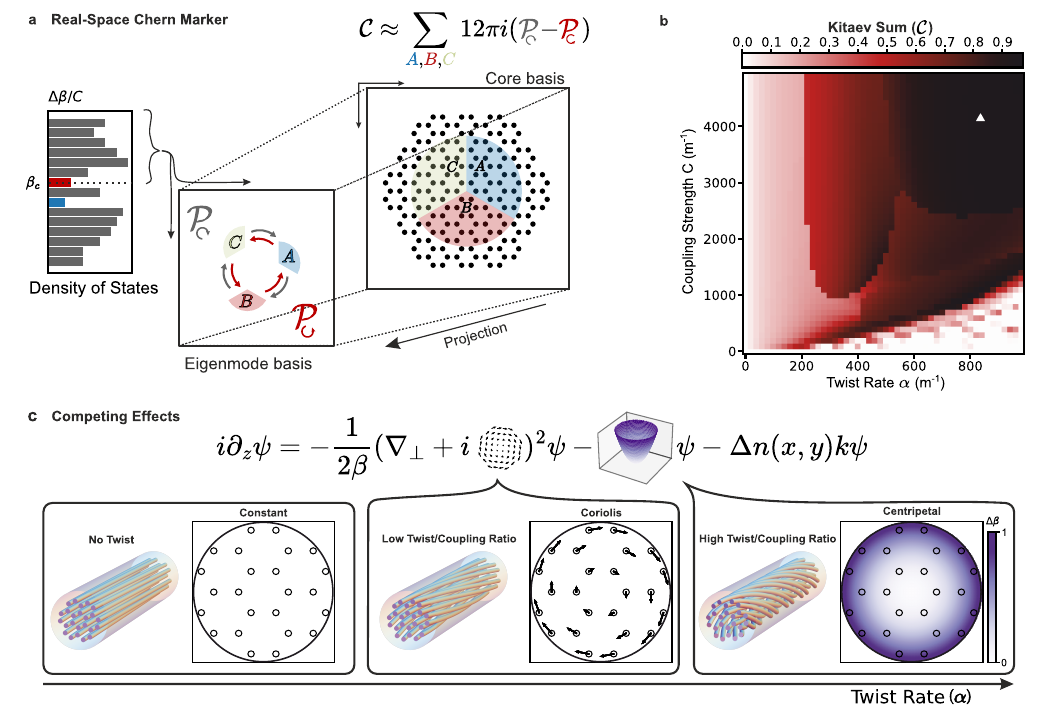}
    \caption{Numerical characterisation of topology in fibre.  \textbf{a}, Schematic explanation of the Kitaev sum used to compute a real-space Chern marker. We define three equal regions  ($A$, $B$, $C$)  and project the eigenmodes into the states above a selected band gap $\beta_c$ using operators ($\mathcal{P}_{A}, \mathcal{P}_{B}, \mathcal{P}_{C}$), respectively.
     The chiral difference $\mathcal{P}_{\circlearrowright} -\mathcal{P}_{\circlearrowleft} = \mathcal{P}_{A} \mathcal{P}_{B} \mathcal{P}_{C} - \mathcal{P}_{C} \mathcal{P}_{B} \mathcal{P}_{A}$ approximates the Chern number in the bulk of our fibre despite inhomogeneity and finite size (see text and SI for details).  \textbf{b}, Calculated real-space Chern marker for different values of twist and coupling strength. We plot experimentally relevant parameters to guide topological fibre design and fabrication. We find the topologically non-trivial region $\mathcal{C} = 1$ corresponds to both high twist rates (greater than \SI{600}{m^{-1}}) and high coupling strength (greater than \SI{3000}{m^{-1}}), and the triangle in the upper right corresponds to our experimentally fabricated fibre. \textbf{c}, Conceptual explanation of the topological transition that occurs as twist rate is increased at a fixed coupling strength, based on Eq.~(\ref{eq:scalarwaveeq}) for guided light in twisted fibre. When the twist is small compared to the coupling strength, the vector-potential (Coriolis) term dominates. Increasing the twist increases the on-site (centripetal) term. At high twist rates, the effect of the centripetal term is to destroy the topology, which we avoid by increasing the coupling strength.}
    \label{fig:3}
\end{figure*}

\begin{figure*}[tp]
    \centering
    \includegraphics[width=\linewidth]{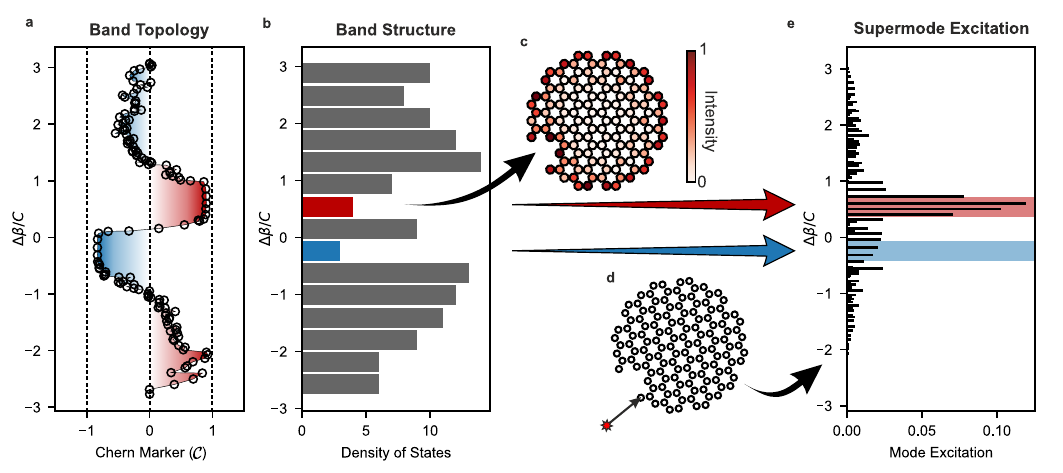}
    \caption{Topological band structure. \textbf{a}, Local Chern marker as a function of rescaled cutoff propagation constant $\Delta \beta/C$. This band structure resolves the Chern marker for bulk states and for each band gap. The two plateaus at $\mathcal{C} \approx \pm 1$ correspond to the two gaps surrounding the 0$^\textrm{th}$ Landau level of a Dirac material, with equal and opposite Chern numbers. \textbf{b}, Histogram showing the density of states as a function of the propagation constant. Topological modes in the band gap regions are highlighted in red and blue, while bulk modes are shown in grey. The topological edge modes live in the two bulk band gaps of the density of states.
    \textbf{c}, Fibre schematic showing a supermode from the upper topological band gap  ($\mathcal{C}  = 1$, red), connecting the presence of robust edge modes with the band topology. \textbf{d}, Diagram showing the core into which light is injected in experiments and finite-element simulations.   \textbf{e}, Overlap integrals between the supermodes (for each $\Delta \beta$)  and the core highlighted in \textbf{d}. Injecting into an edge core predominantly excites the topological states, with the mode in the upper band gap ($\mathcal{C}  = 1$, red) excited more than the counter-propagating mode ($\mathcal{C}  = -1$, blue).}
    \label{fig:4}
\end{figure*}

Experimentally, we probe the topological edge by injecting \SI{1064}{nm} light into a single core at the perimeter of our fibre. In Figures~\ref{fig:1}\textbf{d} and \ref{fig:2}\textbf{a},\textbf{d} we show the intensity distribution after coupling light into the marked core and propagating through \SI{24}{mm} of fibre. Whereas Fig.~\ref{fig:2}\textbf{a} shows a normalised intensity profile, Fig.~\ref{fig:2}\textbf{d} uses a discretised colourmap to highlight the intensity in the darker regions. As explained in the Materials and Methods, our fibre has refractive index $n_{\textrm{SiO}_2} \approx 1.449$ in the silica glass cladding, and $n_{\textrm{doped}} = n_{\textrm{SiO}_2} + \SI[separate-uncertainty = true]{23.0 \pm 0.15}{\times10^{-3}}$ in the Germanium-doped cores (at experimental wavelength). We use a geometry with a core-to-core coupling strength of \SI{4135}{m^{-1}}, twist rate \SI{837}{rad\, m^{-1}}, and core-to-core distance \SI{3.82}{\micro\metre}.
We compare our experimental observations with simulated intensity distributions for our model fibre in the topological ($\alpha =\SI{837}{m^{-1}}$) and trivial ($\alpha = \SI{0}{m^{-1}}$) cases, calculated for the fabricated fibre parameters using finite-element analysis. Simulations of topological fibre in Fig.~\ref{fig:2}\textbf{b} reproduce all of the features of the experimental data, in contrast to the topologically trivial model (Fig.~\ref{fig:2}\textbf{e}), which does not exhibit the same chiral behaviour. 

In simulations, we decompose the propagation of light in fibre into supermodes defined by an unchanging transverse profile.
In Figs.~\ref{fig:2}\textbf{c} and~\ref{fig:2}\textbf{f}, we plot a supermode that demonstrates the edge localisation of propagating light due to the fibre's topology. Figure~\ref{fig:2}\textbf{c} shows an edge mode calculated using  finite-element simulation, while Fig.~\ref{fig:2}\textbf{f} shows the intensity and phase of an edge mode observed in the tight-binding model based on Eq.~(\ref{eq:peierlseqn}), as detailed in SI.
These modes arise in the Chern insulator due to the broken time-reversal symmetry and necessarily follow the arbitrary shape along the perimeter.
To demonstrate this topological classification we now directly calculate a real-space Chern marker. 

\begin{figure*}[t]
\centering
\includegraphics[width=\linewidth]{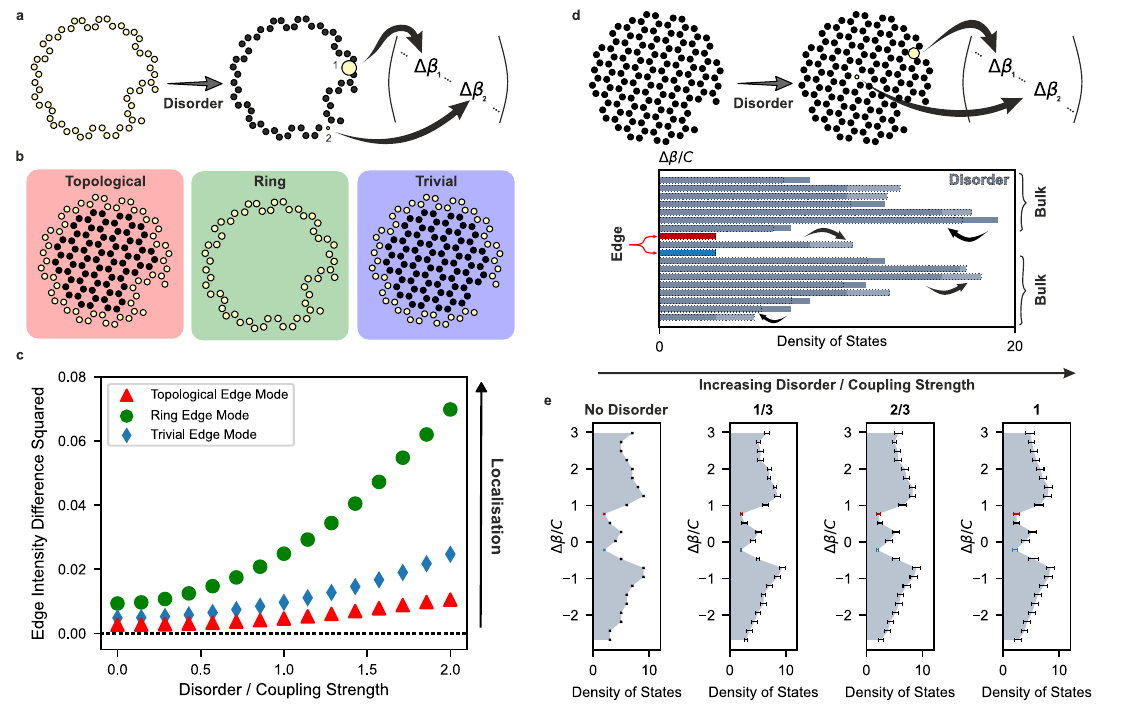}
    \caption{Topological protection against disorder in fibre. \textbf{a}, Schematic of disorder in core shape and size, which can be modelled using random terms on the diagonal elements of the coupling matrix, $C_{jj}$. \textbf{b}, Comparison of three hypothetical fibres in the presence of disorder. Our topological twisted fibre (left, red) has twist rate $\alpha=\SI{837}{m^{-1}}$ and coupling strength $C = \SI{4135}{m^{-1}}$, corresponding to the white triangle in Fig.~\ref{fig:3}\textbf{b}. The ring model (middle, green) is an untwisted array of cores in the same shape as the edge of our topological lattice, which we use to highlight the robustness induced by the topological bulk in our fibre. The trivial model (right, blue) uses a twisted fibre with a large twist, corresponding to a trivial Chern marker (bottom-right of diagram Fig.~\ref{fig:3}\textbf{b}, corresponding to $\alpha=\SI{1700}{m^{-1}}$, $C = \SI{4135}{m^{-1}}$, and  $\mathcal{C} \rightarrow 0$). \textbf{c}, The topological edge mode remains more robust against disorder-induced localisation, for all disorder strengths we consider. We use the Edge Intensity Difference Squared to quantify this localisation, which we compute by squaring and summing the differences between an edge mode in the disordered system and an idealised edge mode that has equal intensity in every edge core. 
    Each curve corresponds to the most delocalised mode in the absence of disorder, and each point is the average over $1000$ disorder realisations.
    \textbf{d}, Schematic illustration for how a single realisation of on-site disorder, introduced across the entire fibre cross-section, changes the density of states for the supported supermodes. \textbf{e}, Numerical data for the density of states for increasing disorder strength (left to right). The error bars correspond to the standard deviation of the density of states across $2000$ realisations of the on-site disorder (see SI for details). Due to topological robustness, the standard deviation is smaller in the band gaps where topological states live (colored red, blue). 
    } 
\label{fig:5}
\end{figure*}

Figure~\ref{fig:3}\textbf{a} gives a diagrammatic explanation of how the Kitaev sum can be used to calculate a local approximation of a system's Chern number~\cite{KITAEV20062}. In this method, the fibre cores are divided up into three equal regions $A,B,C$ in real space. We project the contributions from cores in region $A$ (and $B,C$) into the eigenstates above a selected band gap $\beta_c$ using the operator $\mathcal{P}_{A}$ (and $\mathcal{P}_{B}, \mathcal{P}_{C}$, respectively). To find the local Chern current, we use the difference between the two chiral permutations of the product of the three projection operators, $\mathcal{P}_{\circlearrowright} -\mathcal{P}_{\circlearrowleft} \equiv \mathcal{P}_{A} \mathcal{P}_{B} \mathcal{P}_{C} - \mathcal{P}_{C} \mathcal{P}_{B} \mathcal{P}_{A}$. To find the local Chern marker, we sum over all cores in the regions $A$, $B$, and $C$, see Fig.~\ref{fig:3}\textbf{a}, details in SI, and Ref.~\cite{Mitchell2018}. 

We compute this local Chern marker for parameter values relevant to our fabricated fibre and plot the resulting phase diagram in Fig.~\ref{fig:3}\textbf{b}. The two parameters we experimentally and numerically vary are the inter-core coupling strength (measured in thousands per metre) and the fibre twist rate $\alpha$ (measured in hundreds of radians per metre). We find the local Chern marker $\mathcal{C}$ approaches $\mathcal{C} = 1$ in the region corresponding to both high twist rates (greater than \SI{600}{m^{-1}}) and high coupling strengths (greater than \SI{3000}{m^{-1}}). The white triangle shown in Fig.~\ref{fig:3}\textbf{b}  shows the parameters of the fabricated fibre used in Fig.~\ref{fig:2}, highlighting its location in the Goldilocks zone of the phase diagram.
To understand the interplay between these lengthscales and the topological invariant, we examine the terms in the scalar wave Equation~(\ref{eq:scalarwaveeq}) induced by the fibre twist.

The twist-dependent terms that arise in the scalar wave equation are drawn schematically in Fig.~\ref{fig:3}\textbf{c}.
The breaking of time-reversal symmetry is entirely due to the vector potential term $\mathbf{A} = \alpha \beta (y, -x)$ (Fig.~\ref{fig:3}\textbf{c} middle panel). This term enables a topological band gap due to the photonic equivalent of the Coriolis force when Maxwell's equations are re-expressed in the co-rotating frame of the helicoidal fibre. For small twists, the band gap cannot be observed in our fibre cross-section (or equivalently, the penetration depth of the edge states is larger than fibre cross-section radius $R$). In our experiments, the large values of twist result in a topological band gap of order coupling strength $C$, which also gives the overall bandwidth (setting the fundamental scale for disorder robustness). However, the twist also induces a scalar potential term (Fig.~\ref{fig:3}\textbf{c} right panel), which acts to destroy the topological character of the band structure. This scalar potential is the analogue of a centripetal force term in the co-rotating frame of the fibre, and breaks periodicity for the honeycomb lattice of cores. 
The magnitude of the scalar potential at the edge of the fibre is given by $V(R) = \alpha^2\beta R^2 / 2$, and we expect the topological invariant to break down at large twists when $V(R) > C$. We quantitatively confirm this upper bound on the twist in the SI, showing that at high twists the system instead supports trivial ring-localised modes that lack topological robustness. The topological fibre requires both the twist and the coupling strength to be large, as in the upper-right corner in the Fig.~\ref{fig:3}\textbf{b}, which defines the Goldilocks zone where we have focused our fabrication efforts.

We plot the band structure of the fibre in Fig.~\ref{fig:4}, resolving both the Chern marker and the density of states as functions of the eigenmodes' propagation constant $\Delta \beta$. In Fig.~\ref{fig:4}\textbf{a}, we show that the Chern marker is not well defined in the bulk bands, but shows broad quantised plateaus at $\mathcal{C} = \pm 1$ in the band gaps (plotted in red and blue, respectively), as expected from topological band theory. Figure~\ref{fig:4}\textbf{b} shows that the density of states exhibits two minima for propagation constants corresponding to the topological edge states and the Chern marker plateaus.  Although no wavevector-dependent bands exist due to the non-periodic scalar potential, topological band theory informs how we connect the real-space Chern marker to the edge modes and propagation constants.

To understand the experimental results on chiral edge transport, we numerically simulate injecting monochromatic light into an edge core (see Fig.~\ref{fig:4}\textbf{d}) to excite a superposition of supermodes. 
This superposition can be resolved by computing the overlap integrals for all of the supermodes, plotted in Fig.~\ref{fig:4}\textbf{e}.
Although the supermodes include both bulk and edge states, Fig.~\ref{fig:4}\textbf{e} shows that most of the light is injected into the chiral edge modes (highlighted regions) that live in the topological gaps. 
Crucially, the broad range of propagation constants shown in Fig.~\ref{fig:4}\textbf{e} tells us that both of the topological band gaps are excited simultaneously by this single-core injection.
These two gaps surround the 0$^\textrm{th}$ Landau level and separate it from the $\pm 1$ Landau levels. Analogously to graphene~\cite{Zhang2005, Lado2015}, this band structure  arises due to the flattening of the Dirac cone by the twist-induced vector potential. In our setting, each of the chiral edge modes can be distinguished by its orbital angular momentum (OAM)~\cite{yao2011, xi2013}, which can be computed from the number of nodes along the azimuthal direction in, e.g., Fig.~\ref{fig:2}\textbf{a}.
Due to the notch in our fabricated geometry, we excite the counter-propagating chiral edge states with an overall intensity imbalance, which results in a net chiral transport (see SI for details).

Light transport in chiral edge states is topologically robust. We show two examples of topological robustness that arise in our fibre: Fig.~\ref{fig:5}\textbf{a--c} demonstrates robustness of chiral edge states against disorder-induced localisation, and Fig.~\ref{fig:5}\textbf{d--e} demonstrates robustness of the propagation constants.
In these models, on-site disorder during fibre fabrication changes the size of a core, resulting in a random contribution to the diagonal elements $C_{jj}$ in the coupling matrix, Eq.~(\ref{eq:peierlseqn}) and Fig.~\ref{fig:5}\textbf{a}.

To show topological robustness in Fig.~\ref{fig:5}\textbf{a--c}, we compare the topological fibre (left, red panel in Fig.~\ref{fig:5}\textbf{b}) to two distinct topologically trivial fibres: an untwisted fibre with a ring-like set of cores (middle, green panel in Fig.~\ref{fig:5}\textbf{b}) and a fibre so over-twisted as to be in a topologically trivial regime (right, blue panel in Fig.~\ref{fig:5}\textbf{b}). 
The Edge Intensity Difference Squared plotted in Fig.~\ref{fig:5}\textbf{c} quantifies the modes' robustness against disorder-induced localisation  as a function of disorder strength. We calculate the difference in intensity between a chosen mode and an idealised edge mode that has equal intensity in every edge core, and then square and sum these differences to quantify the localising effect of the disorder. (We provide details in the SI, where we also arrive at the same conclusions for bulk disorder and by measuring localisation via the inverse participation ratio). For each of the topological, ring, and trivial fibres, we select the most edge-localised mode, introduce disorder, and compare how the localisation changes.
In a trivial state, disorder localises modes to only a few cores, unlike the topological state, where the chiral supermode must remain delocalised around the entire edge.

Figure~\ref{fig:5}\textbf{d--e} 
shows that the propagation constants of topological modes are protected against on-site disorder.
The effect of disorder is shown schematically in Figure~\ref{fig:5}\textbf{d}: disorder in one core changes the density of states (per unit $\Delta \beta/C$) for the supermodes. In Fig.~\ref{fig:5}\textbf{e}, we use error bars to plot the standard deviation for the density of states across $2000$ different realisations of on-site disorder, with the disorder strength varying from $0$ to $C$ (see SI for details). We observe that for small disorder strength, the standard deviation in the density of states of the topological edge modes (red and blue) is significantly smaller than the deviations present in the bulk modes (black). We also show that for edge modes, the propagation constant remains topologically protected for a broad range of disorder magnitudes, up to the band-gap size set by nearest-neighbour coupling strength $C$. As expected, once the disorder scale reaches the band-gap size, topological protection is lost and the standard deviation of the density of edge states becomes comparable to the bulk modes. 

In conclusion, we have fabricated a scalable photonic Chern-insulator fibre that supports the propagation of robust, edge-localised supermodes. We have explored the interaction between the twist and the band structure to find a Goldilocks zone where the topological character persists even in the presence of a twist-induced parabolic index profile. We compare our robust states to their trivial counterparts, showing how topology protects light in fibre against fabrication disorder. This robustness underlies the promise of topological fibre design to improve signal reliability, protect delicate quantum signals, and realise scalable topological fibre lasers by doping with gain media.

\subsection*{Funding}
This work is supported by the Air Force Office of Scientific Research under award number FA865522-1-7028. A.S.~acknowledges the support of the Engineering and Physical Sciences Research Council (EPSRC) through New Investigator Award No.~EP/T000961/1 and of the Royal Society under grant No.~RGS/R2/202135.

\clearpage

\begin{widetext}
\renewcommand\thesubsection{\Alph{subsection}}
\setcounter{equation}{0}
\setcounter{section}{0}
\makeatletter
\renewcommand{\theequation}{S\arabic{equation}}
\renewcommand{\thefigure}{S\arabic{figure}}
\setcounter{figure}{0}
\section*{Materials and Methods}
In this section, we outline the numerical methods used to explore the topological properties of our twisted fibre. First, we describe the tight-binding model we use to calculate the local Chern marker in our fibre. Next, we compare the results from our tight-binding model with finite-element-method (FEM) COMSOL simulations. Finally, we explore light propagation in our tight-binding model and COMSOL simulations. 
\section{Tight-Binding Numerical Solution}\label{sec:2}
To find the modes supported by our fibre, we express the solution in terms of the single-core modes that exist in the cross-section. This coupled mode theory approach (also called the tight-binding model) simplifies the numerics and enables us to find our fibre's supermodes from the eigenvalue problem set up in Eq.~(\ref{eq:peierlseqn}). When only considering nearest neighbours, the nonzero terms in the coupling matrix $\mathrm{C}$ for one  mode are given by: 
\begin{align}\label{eq:sum_phi}
\begin{split}
         \Delta \beta_m u_m =&C_{m, m+a_1} e^{i \mathbf{A}\left(\mathbf{m} + \mathbf{a}_1 /2 \right)\cdot \mathbf{a}_1} u_{m+a_1} +  C_{m, m+a_2} e^{i\mathbf{A}(\mathbf{m} + \mathbf{a}_2/2)\cdot \mathbf{a}_2} u_{m+a_2} + \\
         &C_{m, m+a_3} e^{i\mathbf{A}(\mathbf{m} + \mathbf{a}_3/2)\cdot \mathbf{a}_3} u_{m+a_3} + D_{m} u_{m}.
\end{split}
\end{align}
\begin{figure}[bp]
    \centering
    \includegraphics[width=\linewidth]{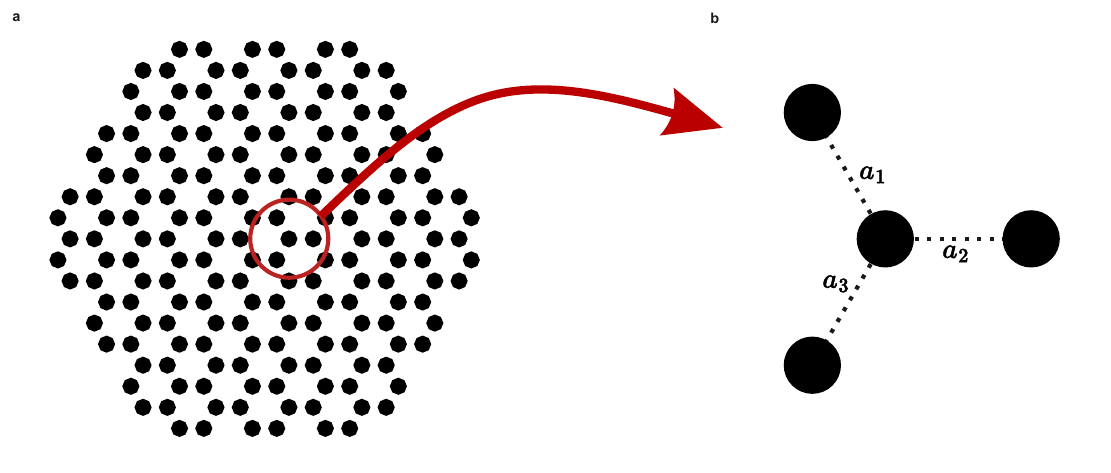}
    \caption{\textbf{a}, Diagram of a 9-ring honeycomb lattice. Each black circle corresponds to a Ge-doped fibre core. \textbf{b}, Couplings between nearest neighbours are labelled with their associated lattice vectors from the central core.}
    \label{fig:s2}
\end{figure}
The first three terms on the right-hand side of Eq.~(\ref{eq:sum_phi}) describe the complex couplings between a chosen core (labeled $m$ and located at $\mathbf{m}$) and its nearest neighbours (shown in Fig.~\ref{fig:s2}). The coupling strength, $C_{m,m+1}$, describes the rate (in units of inverse distance), at which light couples from core $m$ to core $m+1$ when the fibre is untwisted. The coupling strength is governed by the spatial overlap between the supported modes in each core, and is fixed for a given wavelength of light. The Peierls phase is found by solving for the vector potential at the mid-point between each pair of cores and computing the dot product with the lattice vector separating these two cores (corresponding to $\mathbf{a}_{1,2,3}$). The final term in Eq.~(\ref{eq:sum_phi}) describes the on-site radially-dependent change in propagation constant that a core experiences due to the twist. This on-site term is approximated as $D_{m} = \beta_{0} \sqrt{1 + \alpha^2 r_{m}^2} - \beta_{0}$, where $\beta_{0}$ is the propagation for an untwisted core of the same size and shape~\cite{Russell_2017}.

\begin{figure}[tbp]
    \centering
    \includegraphics[width=\linewidth]{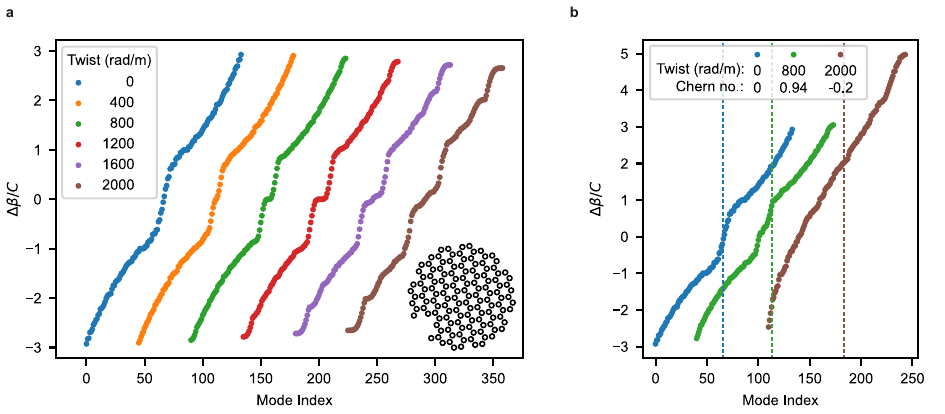}
    \caption{\textbf{a}, Effect of twist rate on the supported propagation constants (calculated using tight-binding numerics). The propagation constants are sorted from lowest to highest eigenvalue $\Delta \beta$ and the mode index corresponds to their position in this order. By plotting the supported propagation constants as a function of mode index, we observe how the band gaps (which correspond to a large spacing between propagation constants) change as a function of the fibre twist rate. As the twist rate increases, the two supported band gaps grow from the original trivial (quasi-)band gap which is present in a Dirac material without twist. Each dataset is shifted horizontally for improved visibility.  \textbf{b}, Propagation constants calculated using the tight-binding numerics, including on-site potential terms, for three values of the twist rate: zero twist, twist below the calculated threshold for over-twisting, and  twist above this  threshold. The legend shows the twist rate and the computed real-space Chern marker.}
    \label{fig:s3}
\end{figure}

We can model an idealised version of our fibre by first neglecting the on-site term $D_m$. Figure~\ref{fig:s3}\textbf{a} shows the computed band structures for our model fibre under this assumption. Varying the twist rate of the fibre changes both the eigenvalues (propagation constants) and eigenmodes (supermodes of the system). First, the trivial band gap closes and two new band gaps appear as Landau levels become better defined. As expected from topological band theory in graphene, these band gaps host robust topological edge modes that enable chiral propagation around the edge of the fibre cross-section. 

Before we consider these states in greater detail, we first add back the on-site potential term that arises in a multicore fibre due to twist. When the on-site terms are introduced along the diagonal of the coupling matrix, if these terms are smaller than the topological band gap (which approaches the coupling strength as the twist is increased), the on-site terms do not change the topological character of the band structure. However, if the on-site terms are greater than the coupling strength (and topological band gap size), the topology breaks down and the system becomes topologically trivial. The coupling strength therefore sets the scale for the topological robustness, and the twist rate at which the on-site terms become greater than the coupling strength forms a natural threshold for topological behaviour. We show in Fig.~\ref{fig:s3}\textbf{b} how the supported propagation constants change above and below this twist rate threshold.

\begin{figure}[tp]
    \centering
    \includegraphics[width=\linewidth]{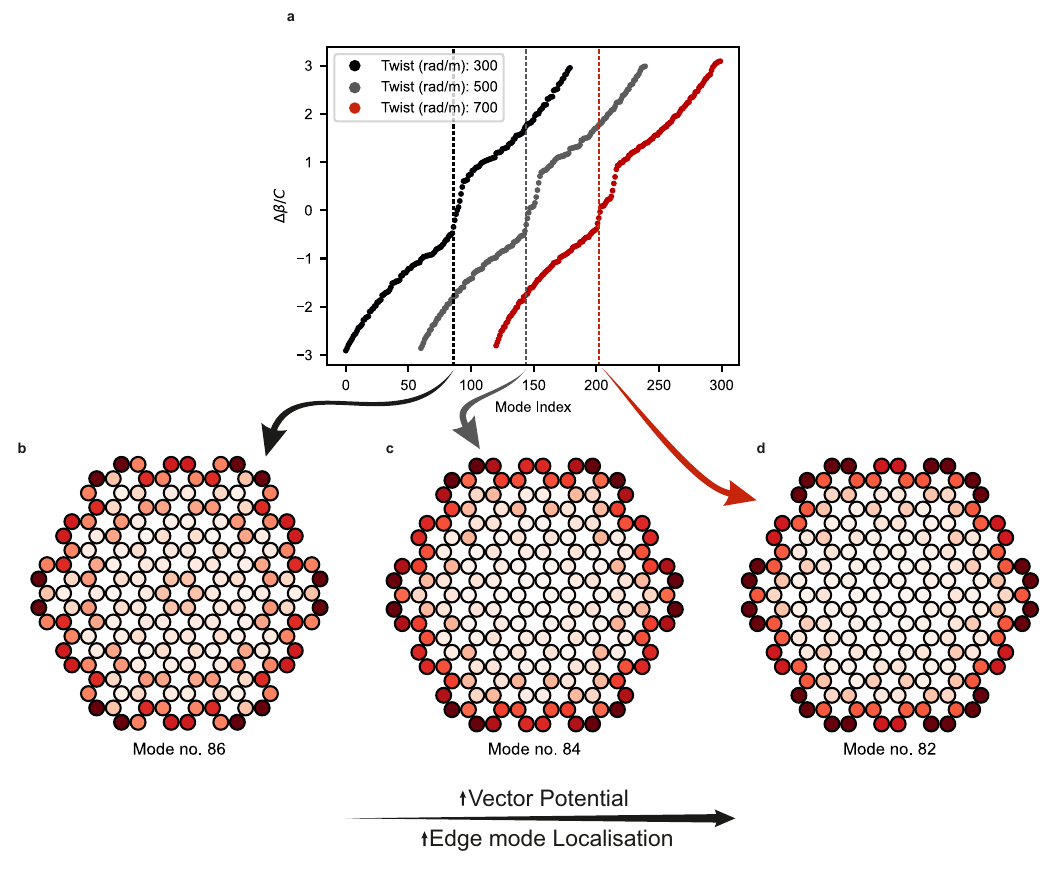}
    \caption{Supported propagation constants and edge mode penetration depth as a function of twist. \textbf{a}, Supported propagation constants for three different fibre twist rates are plotted against their mode index (with a horizontal shift for readability). The mode index labels the supermodes in ascending order (as in Fig.~\ref{fig:s3}). The dotted lines correspond to the same edge mode in each fibre realisation, and their intensity profiles are plotted below.  \textbf{b}, Intensity profile of the selected mode from the smallest-twist band structure in \textbf{a}, for which the edge mode becomes visible. \textbf{c}--\textbf{d}, Intensity profiles of the  band structure corresponding to the middle and largest value of the twist in \textbf{a}, for which the edge mode have smaller penetration depths than in \textbf{b}.}
    \label{fig:s4}
\end{figure}
Figure~\ref{fig:s4} shows the behaviour of the edge modes and band structures below the twist-rate threshold. Figure~\ref{fig:s4}\textbf{a} shows three sets of supported propagation constants for three different values of twist rate (plotted against mode index for visibility). As the twist rate increases, the zeroth Landau level becomes more clearly defined and the band gaps above and below can be more easily identified. By looking at modes in the lower band gap, we can observe the impact of the twist rate on the intensity profiles of the topological supermodes. In Figures~\ref{fig:s4}\textbf{b}-\textbf{d}, we take the same supermode from each dataset and plot the intensity profile. Comparing the intensity profile for each twist rate reveals characteristic changes in the topological edge localisation -- increasing the fibre twist rate increases the edge-localisation of the supported topological supermodes. In a finite lattice, topological edge states have an associated width called the penetration depth~\cite{Lado_2015,Souslov2019}. As we vary the twist (and consequently the vector potential) the penetration depth of the edge modes gets smaller and they become more localised to the cores at the perimeter of the system. To enable clear observation of the topological edge modes, the penetration depth must be sufficiently small, and thus the twist rate sufficiently high. If the penetration depth is large, the topological mode will have a greater overlap with cores in the bulk of the fibre. This greater overlap means that light coupled into the edge of the fibre will not stay robustly localised, and instead light will couple into the bulk of system and the characteristic topological localisation will be obscured. In our system, we balance our requirement of a small penetration depth with the parabolic scalar potential term that we introduce due to the twist.  

Twisting our fibre introduces both on-site (centripetal) and vector potential terms into the scalar wave equation. To create a topologically non-trivial fibre, we must ensure that the topological band gap induced by the vector potential is greater than the on-site (centripetal) term. However, as the twist rate increases, the on-site (centripetal) term grows faster than the band-gap size, leaving the fibre topologically trivial at high twist rates. When twisting a fibre, the supported propagation constants of each core begin to vary as a function of radial distance. As supermodes can only form from cores that have similar propagation constants, once the on-site potential dominates and the fibre becomes topologically trivial, the supermodes form ring shapes at fixed radii. These ring-localised modes are no longer protected by a topological band gap, but are trivially localised into ring shapes. We demonstrate the difference between topological edge states and ring-localised modes in Fig.~\ref{fig:s5}\textbf{a} and Fig.~\ref{fig:s5}\textbf{b} and compare our tight-binding predictions to the finite-element solutions found using COMSOL Multiphysics. 

\section{COMSOL simulations}
To solve for the electromagnetic modes of our twisted fibre cross-section, we combine the cross-sectional geometry of the fibre with the co-twisting coordinate frame. We then introduce a reverse coordinate transform to convert the numerical solutions into the lab frame. The reverse coordinate transform we use is:
\begin{align}
\label{eq:reversecoordtransform}
    x &= x' \cos(\alpha z') + y' \sin(\alpha z'), \nonumber\\
    y &= -x' \sin(\alpha z') + y' \cos(\alpha z'),    \nonumber\\ z & = z',
\end{align}
where ($x,y,z$) are the coordinates in the lab frame, ($x', y', z'$) are the coordinates in the co-twisting frame, and $\alpha$ is the twist rate in radians \SI{}{m^{-1}}.
As highlighted in Ref.~\cite{Nicolet2008}, changing the coordinate frame in which the Maxwell equations are described is equivalent to changing the permittivity and permeability tensors that feature in Maxwell's equations. Instead of solving Maxwell's equations in a helicoidal frame with scalar permeabilities and permittivities, we introduce the effects of twist into the permittivity and permeability tensors and solve for the supermodes in an untwisted geometry. Following the approach in Refs.~\cite{Russell_2017,Nicolet2008}, we replace the permittivity ($\varepsilon$) and permeability ($\mu$) tensors for the untwisted case, with $\varepsilon' = \varepsilon T^{-1}$, $\mu' = \mu T^{-1}$, where $T$ is determined by the Jacobian of the coordinate transform: 
\begin{equation}
    \mathrm{T} = \frac{\mathrm{J}^T \mathrm{J}}{\textrm{det}(\mathrm{J})} = \begin{bmatrix}
    1 &  0 & \alpha y'\\
    0  & 1 & -\alpha x'\\
    \alpha y' & -\alpha x' & 1 + \alpha^2(x'^2 + y'^2) 
    \end{bmatrix}.
\end{equation}
\begin{figure}[hp]
    \centering
    \includegraphics[width=\linewidth]{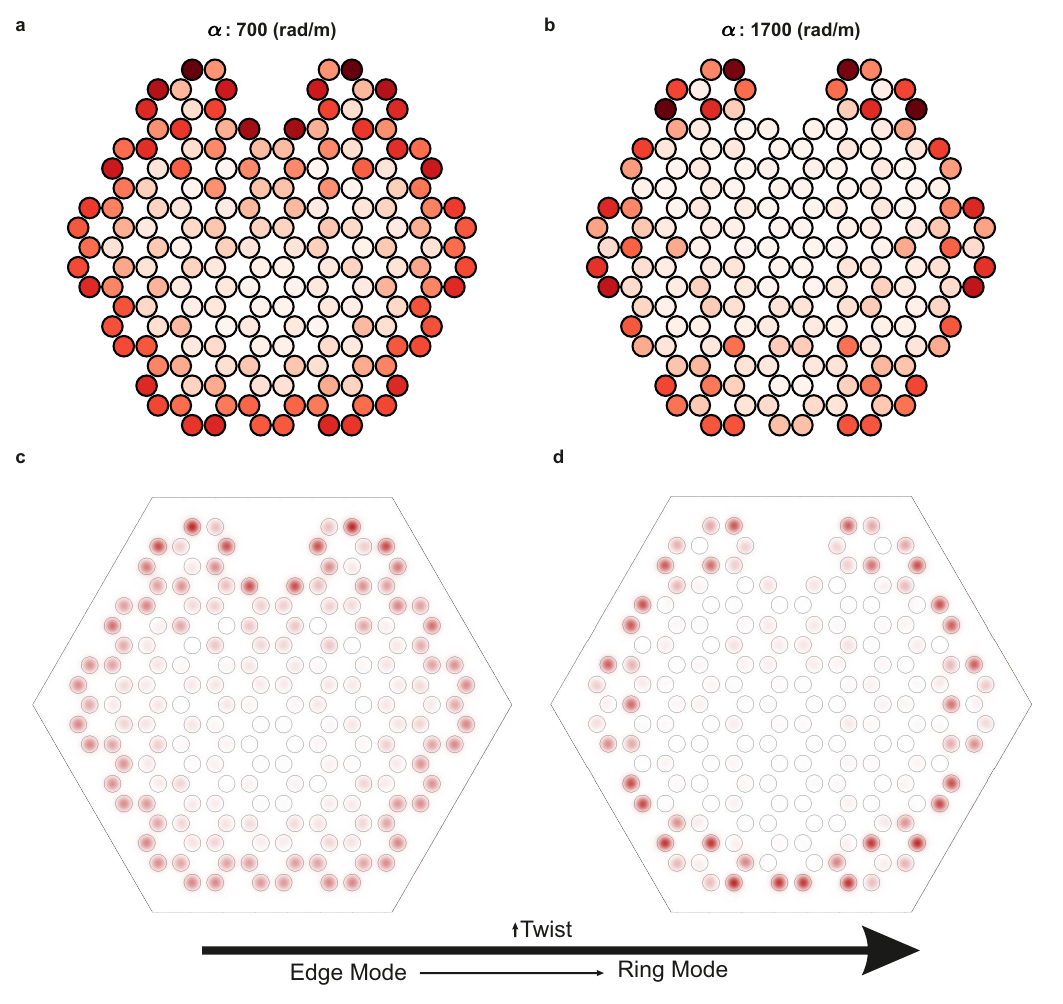}
    \caption{Intensity profiles above and below the topological-to-trivial twist-rate threshold. \textbf{a}, Edge mode intensity profile calculated using tight-binding numerics for a fibre with a twist rate of \SI{700}{\radian \per  \m}. The \SI{700}{\radian \per  \m} twist rate introduces an on-site term that is smaller than (but of the same order as) the coupling strength, so the topological edge-localisation remains. \textbf{b}, Intensity profile for a ring-localised mode calculated at a fibre twist rate of \SI{1700}{\radian \per  \m}. A twist rate of \SI{1700}{\radian \per  \m} introduces an on-site term which is greater than the coupling strength (which defines the topological band gap), so the topological state is destroyed and trivial ring-localised supermodes remain. \textbf{c--d}, Corresponding data from finite-element simulations: \textbf{c}, Intensity profile for an edge mode calculated using finite-element simulation for a fibre with a twist rate of \SI{700}{\radian \per  \m}.\textbf{d}, Intensity profile for a ring-localised mode calculated in a finite-element simulation of fibre with a twist rate of \SI{1700}{\radian \per  \m}.
    }
    \label{fig:s5}
\end{figure}
\begin{figure}[tp]
    \centering
    \includegraphics[width=\linewidth]{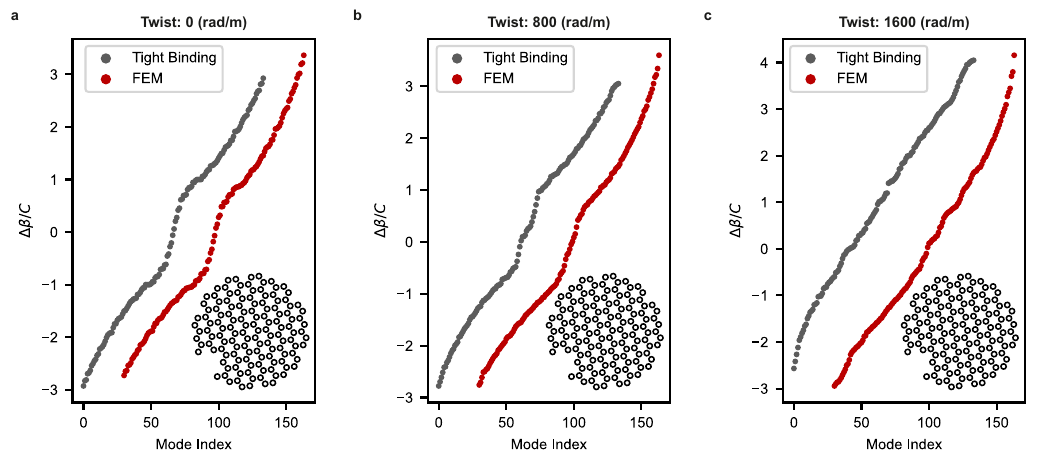}
    \caption{Comparison between tight-binding numerics and finite-element simulation shows good quantitative agreement between these distinct computational methods across a range of twist rate values.
    \textbf{a}, Propagation constants calculated using tight-binding numerics and finite-element simulation with no twist (i.e., a twist rate of \SI{0}{\radian \per  \m}) plotted against their mode index to highlight their similarities and differences. The mode indices are found by sorting the propagation constants in ascending order, and we use them here to allow for a clearer comparison between datasets. The finite element method datasets are shifted in mode index to increase visibility and prevent overlap. \textbf{b}, Propagation constants calculated using tight-binding numerics and finite-element simulation with a twist rate of \SI{800}{\radian \per  \m}, plotted against the mode index. \textbf{c}, Propagation constants calculated using tight-binding numerics and finite-element simulation with a twist rate of \SI{1600}{\radian \per  \m}, plotted against mode index. All propagation constants are calculated for the fibre cross sections shown in the insets.}
    \label{fig:s6}
\end{figure}

\begin{figure}[tbp]
    \centering
    \includegraphics[width=\linewidth]{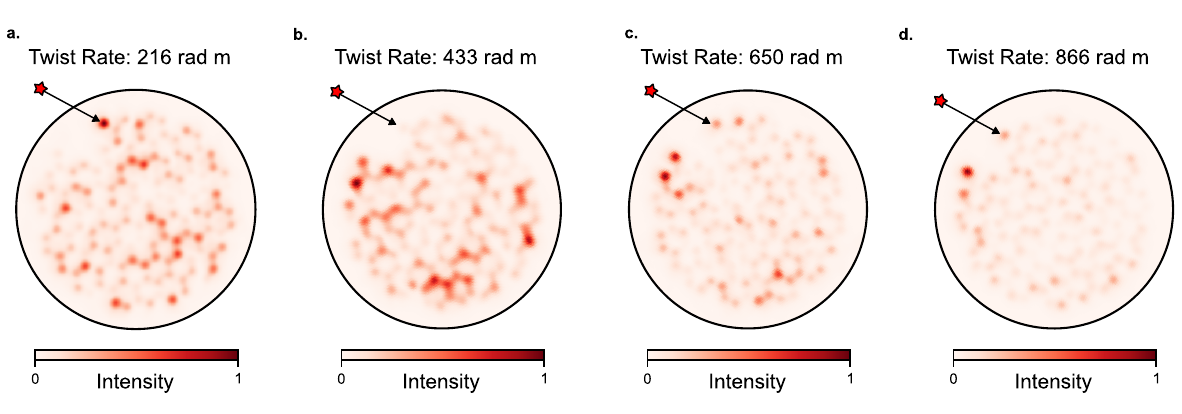}
    \caption{Varying twist rate in experimental results. Light propagates through four different fibres, fabricated with different twist rates but the same cross-sectional structure. Light is injected into the marked core and propagates through \SI[separate-uncertainty = true]{23\pm2}{mm} of each fibre. We see increasing localisation of intensity as a function of the fibre twist rate. \textbf{a}, Light is injected into a lightly twisted fibre (\SI{216}{m^{-1}}) and is not localised to the perimeter. \textbf{b}, Light propagates through a fibre with a twist rate \SI{433}{m^{-1}} and remains delocalised at the output. \textbf{c}, Intensity becomes more localised at the edge of the lattice when light propagates through a fibre with twist rate \SI{650}{m^{-1}}. \textbf{d}, Edge-localisation of intensity can be seen when light propagates through a fibre with twist rate of \SI{866}{m^{-1}}.}
    \label{fig:s11}
\end{figure}

After replacing the permittivity and permeability tensors, we can solve for the supermodes of the fibre cross-section. Two of these supermodes, at different levels of twist, are shown in Fig.~\ref{fig:s5}\textbf{c},\textbf{d}. Figure~\ref{fig:s5}\textbf{c} shows a topological edge mode for a 9-ring honeycomb lattice with a single hexagon missing. The FEM (Finite Element Method) solution is in good agreement with our tight-binding numerics, and shows a mode that remains localised to the perimeter, even in the presence of a missing hexagon. As explained in section~\ref{sec:2} and the main text, when we increase the twist rate beyond the topologically protected threshold (i.e., when the on-site scalar potential term introduced by the twist is greater than the band gap size), the edge mode is destroyed and only ring-localised modes remain. Figures~\ref{fig:s5}\textbf{b} and ~\ref{fig:s5}\textbf{d} show a ring-localised mode arising from our solutions. Figure~\ref{fig:s5}\textbf{b} shows the tight-binding prediction in good agreement with the finite-element simulation Fig.~\ref{fig:s5}\textbf{d}. In contrast with the topological model shown in Fig.~\ref{fig:s5}\textbf{a} and Fig.~\ref{fig:s5}\textbf{c}, the intensity in the ring-localised mode is not uniform over the whole perimeter of the fibre. The cut-out section at the top of the lattice does not maintain a uniform intensity in the ring-localised cases (Fig.~\ref{fig:s5}\textbf{b} and Fig.~\ref{fig:s5}\textbf{d}) but in the topological cases Fig.~\ref{fig:s5}\textbf{a} and Fig.~\ref{fig:s5}\textbf{c}, intensity is localised evenly across all cores at the edge of the fibre. 

We can probe the band structure of our FEM system by solving for the supported propagation constants of our twisted fibre cross section (see Fig.~\ref{fig:s6} and, for comparison, experimental results in Fig.~\ref{fig:s11}). Figures~\ref{fig:s6}\textbf{a}, ~\ref{fig:s6}\textbf{b}, and ~\ref{fig:s6}\textbf{c} show the comparison between supported propagation constants calculated using finite-element and tight-binding numerics. When there is no twist, both the FEM and tight-binding model show only a single visible bandgap (see Fig.~\ref{fig:s6}\textbf{a}). Below the twist disorder threshold, both the tight binding and COMSOL solutions show two band gaps, punctuated with a small band of modes at zero change in propagation constant (see Fig.~\ref{fig:s6}\textbf{b}). Once the twist-induced disorder is greater than the band-gap size, the topological band gap and associated band structure are lost in both FEM and tight binding solutions (see Fig.~\ref{fig:s6}\textbf{c}). 

\section{Numerical Propagation}
To simulate propagation in the tight-binding numerics, we take an initial field corresponding to a single core excitation (of the marked core in Figs.~\ref{fig:s9}\textbf{b} and ~\ref{fig:s9}\textbf{c}) and by using overlap integrals, express this excitation as a sum of supermodes. These supermodes propagate unchanged in $x$- and $y$-coordinates, with the phase evolving as $e^{i\beta_{s} z}$, where $\beta_{s}$ is the propagation constant of each supermode. At each step of propagation, we map the intensity into the tight-binding basis, and calculate the resultant intensity profile. A similar process is used for the COMSOL solutions, with the significant difference that both the supermodes and the initial excitation are now expressed on a $100\times100$ grid that spans the fibre cross-section. By discretising the supermodes into this $100\times100$ basis, we gain more information about the spatial structure of our solutions while accurately scaling to arbitrary propagation distances. 

Using the computed intensity profiles, we then calculate how the distribution of intensity changes as light propagates along the fibre (see Fig.~\ref{fig:s9}\textbf{a}). As shown in Fig.~3 of the main text and in Fig.~\ref{fig:s9}, when light is injected into a single core, a range of supermodes is excited. If these excited supermodes feature an uneven distribution of light within each topological band gap, the resultant intensity profile will move around the edge of the system and have an associated chirality. This chirality arises due to the presence of the one-way edge states that are guaranteed by the presence of the Chern topological invariant. In our system, the upper and lower band gaps feature equal and opposite Chern numbers, giving rise to two opposite chiral edge currents. In a system like graphene in a magnetic field, one can set the Fermi level such that only a single chiral edge state can be accessed~\cite{Lado_2015}, but in our optical fibre we cannot set an effective Fermi level. Instead, we necessarily excite superpositions of states from both band gaps. If we excite an equal superposition of these Chern states, the overall propagation of the intensity profile does not have a chiral character. However, if there is greater overlap with modes of one chirality, we observe chiral propagation. 

To demonstrate this chirality, we find the centre of intensity for every step along the fibre and plot its angle relative to the origin (see Fig.~\ref{fig:s9}\textbf{a}). In Fig.~\ref{fig:s9}\textbf{b} we compare the predicted change in $\theta$ (the angle made to the centre of intensity from the lattice origin) as a function of distance for tight-binding and finite-element propagation. Both the tight-binding and finite-element solution describe a change in the centre of intensity with a single average direction (increasing $\theta$, which corresponds to an anticlockwise movement of light). This indicates that the unequal excitation of topological edge modes gives rise to chiral transport of light intensity within the fibre cross-section. In Fig.~\ref{fig:s9}\textbf{c}, we compare this movement of the centre of light intensity with the numerical predictions of an untwisted fibre over the same length to show that the chirality only arises in the presence of our effective magnetic field. In the trivial case, the angle $\theta$ no longer moves in one consistent direction.
\begin{figure}[t]
    \centering
    \includegraphics[width=\linewidth]{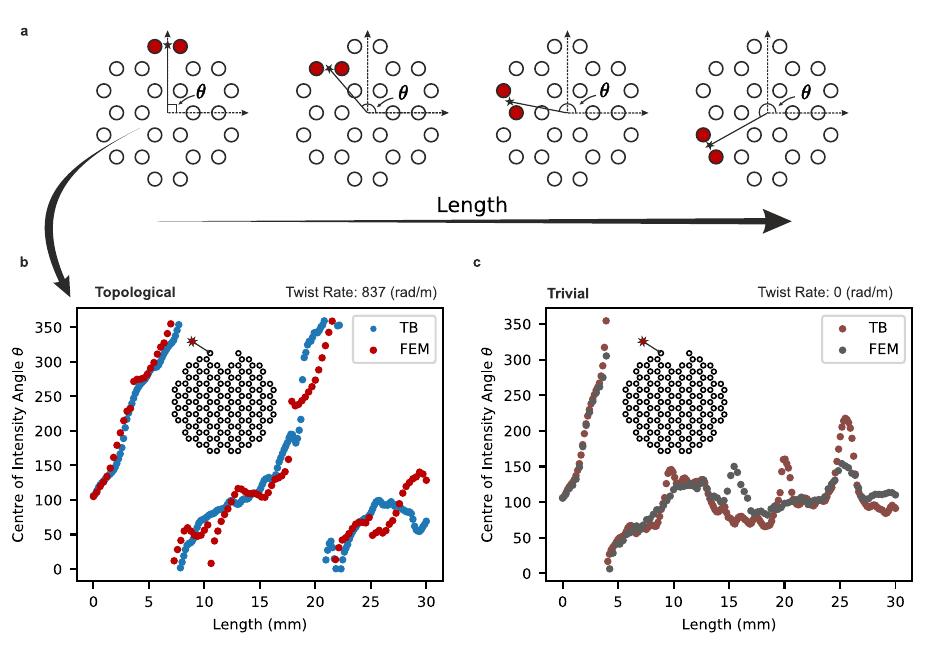}
    \caption{Chiral transport of light intensity. \textbf{a}, Diagrammatic explanation of how the centre of intensity ($\star$) changes as a function of distance. The angle this centre of intensity makes to the origin, $\theta$, is used to determine the chiral character of transverse light propagation. \textbf{b}, The angle that the centre of intensity makes to the origin, $\theta$, is plotted as a function of propagation length for tight-binding (blue) and finite-element (red) models based on our fabricated fibre parameters. Light is initially injected into the core highlighted in the inset. \textbf{c}, The angle that the centre of intensity makes to the origin, $\theta$, is plotted as a function of propagation length for tight-binding (brown) and finite-element (grey) models based on an untwisted version of our fabricated fibre, where no consistent chiral propagation is observed. Light is initially injected into the core highlighted in the inset.}
    \label{fig:s9}
\end{figure}

\section*{Supplementary Material}
In this section, we present detailed derivations of the vector and scalar wave equations that describe light propagation in twisted fibre. In addition, we show that twisting the fibre breaks propagation symmetry for an optical mode in a way closely analogous to how an external magnetic field breaks time-reversal symmetry for an electron. Finally, we explore the topological nature of our tight-binding model in greater depth. We show how the measured real-space Chern marker changes as a function of both the vector and scalar potentials, and describe in detail how we can assess the disorder robustness of our supported topological states. 
\setcounter{section}{0}
\section{Fibre Twist as a Vector Potential}
Here we adapt the derivation of the paraxial approximation from the Ref.~\cite{Rechtsman2013}, with the significant difference that in our fibre, the cross-section rotates as a rigid body along the propagation direction.
To go to the co-twisting frame, our coordinate transform describes all of the fibre cores twisting helically around a single origin. By contrast, Ref.~\cite{Rechtsman2013} considers the individual twist of each core---a geometry that is inaccessible in optical fibre, even with state-of-the-art fabrication methods. At the end of the derivation, we arrive at a Schr\"{o}dinger-like equation [Eq.~(4) of the main text]  in the presence of a vector potential, where the vector potential is in the symmetric gauge. 

We use a combination of analytical and numerical approaches to solve this Schr\"{o}dinger-like equation in an external gauge field. First, we consider a tight-binding model and convert the vector potential using a Peierls substitution, in order to numerically solve for the supermodes. Our second approach is to use finite-element COMSOL simulations to calculate the supermodes of the fibre cross-section.  
\subsection{ASSUMPTIONS} 
To begin the derivation, we make the following assumptions, which apply to light propagation in fibre: 

\begin{itemize}
    \item The relative permeability $\mu_{r}$ at optical frequencies is given by $\mu_{r} = 1 $.
    \item The divergence of the electric field $\mathbf{E}$ vanishes, i.e., $\nabla\cdot \mathbf{E} = -\frac{1}{n^2} \mathbf{E} \cdot \nabla n^2 = -\nabla(\mathbf{E}\cdot \nabla \ln n^2) = 0$. This is because the field is small wherever the change in the refractive index, $\nabla \ln{n^2}$, is non-zero.  
    \item The envelope of the electric field varies much slower than a period of oscillations [this is called the slowly-varying envelope approximation (SVEA)]: $ |\partial_{z}^2 \mathbf{\Psi}| \ll  \beta |\partial_{z} \mathbf{\Psi}|$.
    \item The field solution, $\mathbf{E}(x,y,z,t)$ separates into, e.g.,   $\psi(x,y,z)e^{i(\beta z -\omega t)} \hat{\mathbf{x}}$.
    \item The change in refractive index is small (known as the weak guidance approximation), so we can expand the square: $n^2 - n_{0}^2 = (n + n_{0})(n - n_{0}) \approx 2 n_{0} (\Delta n)$.
    \item Due to the small change in index, the propagation constant can be approximated using $\beta \approx k n_{0}$, where $k$ is the free-space wavenumber.

\end{itemize}
\subsection{VECTOR WAVE EQUATION}
Light propagating in a waveguide of arbitary index $n = n(x,y,z)$ is described by the Maxwell equations:
\begin{align}\label{eq:Maxwell}
    &\nabla\cdot(n^2\mathbf E)=0, && 
    \nabla\times\mathbf H = \epsilon_0 n^2 \partial_t\mathbf{E},\\
    & \nabla\cdot\mathbf H = 0, && \nabla\times\mathbf E = -\mu_0\partial_t\mathbf{H}.
\end{align}
The vector wave equation is obtained using the common curl-curl identity and assuming $\nabla\cdot\mathbf{E}=0$:
\begin{equation}\label{eq:curlcurleqn}
    \nabla^2 \mathbf{E} =  \mu_{0} \varepsilon_{0} n^2 \partial_{t}^{2} \mathbf{E}.
\end{equation}
We then use the plane-wave solutions, $\mathbf{E}(x,y,z,t) = \mathbf{\Psi}(x,y,z)e^{i(\beta z - \omega t)}$, to write  the vector Helmholtz equation:
\begin{align}
    \nabla^2[\mathbf{\Psi} e^{i(\beta z - \omega t)}] &=
    \mu_{0} \varepsilon_{0} n^2 \partial_{t}^{2}[\mathbf{\Psi} e^{i(\beta z - \omega t)}]\label{eq:scalarattempt1}\\
    &= -\omega^2\mu_0\varepsilon_0 n^2 [\mathbf{\Psi} e^{i(\beta z -\omega t)}],\\
    \implies (\nabla^2 + k^2 n^2)&[\mathbf{\Psi} e^{i(\beta z - \omega t)}] = 0.\label{eq:Helmholtz}
\end{align}
with Eq.~(\ref{eq:Helmholtz}) making use of $\mu_0\varepsilon_0 = \frac{1}{c^2}$ and $k = \frac{\omega}{c}$. 

We now make the coordinate transformation into the helicoidal frame, in which the three-dimensional refractive index profile $n(x,y,z)$ of the twisted fibre is transformed into an effective two-dimensional refractive index profile $n(x,y)$.
In order to transform into the helicoidal frame, we rotate the electric field into the co-rotating frame via
\begin{align}
    \mathbf{E}(\mathbf{v})\mapsto R(\tau z)\cdot\mathbf{E}(R^T(\alpha z)\cdot\mathbf{v}),
\end{align}
where $\mathbf{v} = \{x,y,z\}$, and $R$ is the rotation matrix of the helicoidal transformation. We show this rotation schematically in Fig.~\ref{fig:s1}\textbf{a}. In this notation, $R(\tau z)$ and $R(\alpha z)$ are the helicoidal rotation matrices with arguments $\tau z$ and $\alpha z$, respectively,  where $\tau = \frac{\alpha}{1+\alpha^2 r^2}$ is the torsion of the helical path and $\alpha$ is the twist rate of the fibre. The two rotation matrices correspond to different rotation angles because the coordinates rotate with the twist rate of the fibre, i.e., by $\alpha$, whereas, the electric field rotates by the torsion $\tau$~\cite{ross}. Significantly, for our fibre, the ratio $\alpha/\tau$ is close to 1. Explicitly, these transformations take the form:
\begin{align}
\partial_z \mathbf{E} \mapsto \partial_{z}\mathbf{E} + \boldsymbol{\tau}\times\mathbf{E}
\label{eq:VWEtransformations1}
\end{align}
with 
\begin{align}
 \begin{pmatrix}
    x \\ y \\ z\end{pmatrix} = \underbrace{\begin{pmatrix}
        c' & s' & 0 \\ -s' & c' & 0 \\ 0 & 0 & 1
    \end{pmatrix}}_{R(\alpha z')}
    \begin{pmatrix}
        x'\\y'\\z'
    \end{pmatrix},
&& \begin{pmatrix}
        \partial_x \\ \partial_y \\ \partial_z
    \end{pmatrix}
    =
    \underbrace{\begin{pmatrix}
        c' & s' & 0 \\
        -s' & c' & 0 \\
        -\alpha y' & \alpha x' & 1
    \end{pmatrix}}_{J^{-T}(\alpha z')}
    \begin{pmatrix}
        \partial_{x'} \\ \partial_{y'} \\ \partial_{z'}
    \end{pmatrix},\label{eq:VWEtransformations2}
\end{align}
where $\boldsymbol{\tau} = \tau\hat{z}$ is the angular rotation rate of the electric field in $\SI{}{rad\,m^{-1}}$, and $c' = \cos{\alpha z'}$ (similarly, $s' = \sin{\alpha z'}$). The inverse transpose Jacobian, $J^{-T}$, is derived from the partial derivatives of $R$, and describes the coordinate transformation of the derivatives. Substituting the electric field from Eq.~(\ref{eq:VWEtransformations1}) into Eq.~(\ref{eq:Helmholtz}) introduces additional terms into the effective Hamiltonian that mix the transverse components of the field,
\begin{align}
    (\nabla^2 + k^2n^2)[\mathbf{\Psi} e^{i(\beta z - \omega t)}] + 2i\beta\boldsymbol{\tau}\times[\mathbf{\Psi} e^{i(\beta z - \omega t)}]&= 0.\label{eq:tauVWE}
\end{align}
Eq.~(\ref{eq:tauVWE}) contains a second axial derivative of the field, $\partial_{z'}^2[\mathbf{\Psi}e^{i(\beta z - \omega t)}]$, which we  expand using the product rule,
\begin{align}\label{eq:expansion}
     \partial_z^2[\mathbf{\Psi} e^{i(\beta z - \omega t)}] = 
     - \beta^2 \mathbf{\Psi} e^{i(\beta z - \omega t)} + e^{i(\beta z - \omega t)} \partial_{z}^{2} \mathbf{\Psi} + 2i \beta e^{i(\beta z - \omega t)} \partial_{z} \mathbf{\Psi} .
\end{align}
We neglect the second derivative of the field amplitude, $\partial_z^2\mathbf{\Psi}$, using the slowly-varying envelope approximation, $ \vert\partial_{z}^2 \mathbf{\Psi}\vert \ll \beta\vert\partial_{z} \mathbf{\Psi}\vert$. Substituting the expansion Eq.~(\ref{eq:expansion}) into Eq.~(\ref{eq:tauVWE}) yields,
\begin{align}
    \nabla_\perp^2\mathbf{\Psi} + 2i\beta(\partial_z \mathbf{\Psi} + \boldsymbol{\tau}\times\mathbf{\Psi}) + (k^2 n^2 - \beta^2)\mathbf{\Psi} = 0.\label{Eq:SVEA2}
\end{align}
Here, each term has been divided by the modulating phasor, $e^{i(\beta z - \omega t)}$. Using $\beta \approx k n_{0}$, we obtain
\begin{equation}
    \nabla_{\perp}^2 \mathbf{\Psi} + 2i k n_{0}(\partial_{z} \mathbf{\Psi} + \boldsymbol{\tau}\times\mathbf{\Psi}) + k^2( n^2 - n_{0}^2) \mathbf{\Psi} = 0.
\end{equation}
We then approximate $n^2 - n_{0}^2 = (n + n_{0})(n - n_{0}) \approx 2 n_{0} \Delta n(x,y,z)$, to find
\begin{equation}
    \nabla_{\perp}^2 \mathbf{\Psi} + 2i k n_{0}(\partial_{z} \mathbf{\Psi} + \boldsymbol{\tau}\times\mathbf{\Psi}) + 2 k^2 n_{0} (\Delta n) \mathbf{\Psi} = 0.
\end{equation}
Splitting the terms so that the $z$-derivative is on the right-hand side, we arrive at the analogue of the Schr\"{o}dinger wave equation,
\begin{align}
\label{eq:scalarwave}
     i \partial_{z} \mathbf{\Psi} = -\frac{1}{2 \beta} \nabla_{\perp}^2 \mathbf{\Psi} - k(\Delta n) \mathbf{\Psi} -i \boldsymbol{\tau}\times\mathbf{\Psi}.
\end{align}
\begin{figure}[tp]
    \centering
    \includegraphics[width=\linewidth]{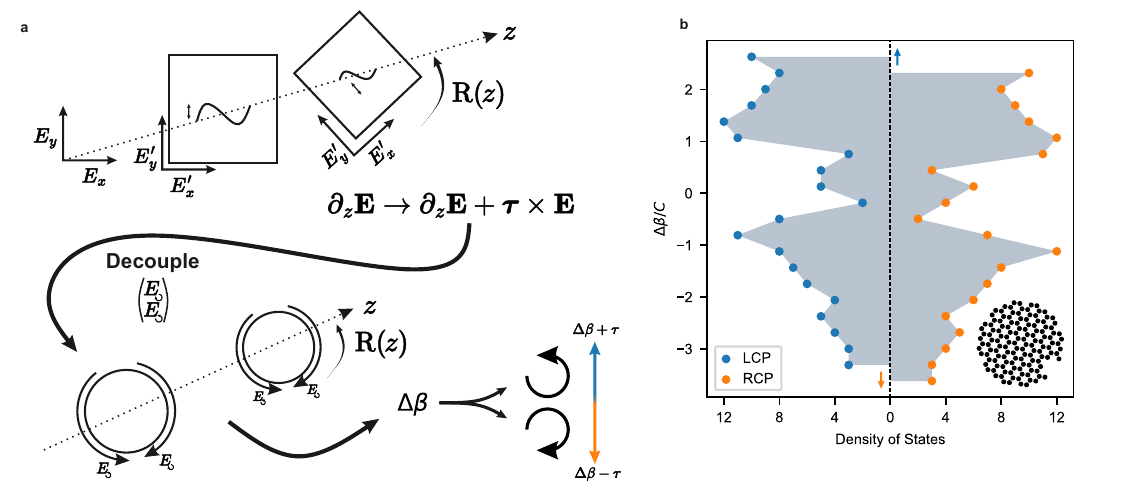}
    \caption{Polarisation effects in twisted optical fibre. \textbf{a}, Twisting the fibre leads to a rotation of the electric field along the $z$-axis. Moving into a co-twisting frame reveals a mixing in the $E_{x}$ and $E_{y}$ field components. We can return to two uncoupled equations by describing the electric field in the circular polarised basis---where each polarisation is shifted by $\pm\boldsymbol{\tau}$. \textbf{b}, Finite element simulation is used to calculate supermode data which is projected into left and right circularly polarised bases. Each polarisation has the same general structure in its density of states, with an overall shifting of $\pm\boldsymbol{\tau}$.}
    \label{fig:s1}
\end{figure}

We now perform the coordinate transform by substituting equations Eq.~(\ref{eq:VWEtransformations2}) into Eq.~(\ref{eq:scalarwave}). Noticing that the transverse Laplacian is rotation invariant, i.e. $\nabla_{\perp}^2 \mathbf{\Psi} = \nabla_{\perp}^{'2} \mathbf{\Psi}'$, we have,
\begin{align}\label{eq:angular}
    i\partial_{z'}\mathbf{\Psi}' =-\frac{1}{2\beta}\nabla_\perp'^2 \mathbf{\Psi}' - \alpha\hat{L}_{z'}\mathbf{\Psi}' - k(\Delta n')\mathbf{\Psi}' -i\boldsymbol{\tau}\times \mathbf{\Psi}',
\end{align}
where $\hat{L}_{z'} = -i(x'\partial_{y'} - y'\partial_{x'})$ is the $z'$ component of the angular momentum operator. These terms arise due to the action of the Jacobian on  the $z$-derivative in Eq.~(\ref{eq:VWEtransformations2}):
\begin{align}
    \partial_{z} = \partial_{z'} - \alpha y' \partial_{x'} + \alpha x'\partial_{y'}. 
\end{align}

We rewrite two of the terms (the Laplacian and the angular momentum terms) from Eq.~(\ref{eq:angular}), by completing the square, to obtain terms of the form of a vector potential and a parabolic scalar potential, i.e.,
\begin{align}\label{eq:vectorpotential}
     -\frac{1}{2\beta} \nabla'^2_{\perp} \mathbf{\Psi}' -\alpha\hat{L}_{z'}\mathbf{\Psi}' &= -\frac{1}{2 \beta} (\nabla'_{\perp} + i\mathbf{A})^2\mathbf{\Psi}' - \frac{\alpha^2\beta r'^2}{2}\mathbf{\Psi}',
\end{align}
where we define $ \mathbf{A} \equiv \alpha \beta (y', -x')$ and $r'^2 = x'^2 + y'^2$. 
Substituting Eq.~(\ref{eq:vectorpotential}) into Eq.~(\ref{eq:angular}), we obtain
\begin{align}\label{vectorwaveeq}
    i\partial_{z'} \mathbf{\Psi}' &= -\frac{1}{2\beta}(\nabla' + i \mathbf{A})^{2} \mathbf{\Psi}' - \frac{\alpha^2 \beta r'^2 }{2} \mathbf{\Psi}' - k \Delta n' \mathbf{\Psi}' -i \boldsymbol{\tau}\times\mathbf{\Psi}'\\
    \mathbf{A} &= \alpha \beta (y', -x')\nonumber
\end{align}
To treat the polarisation degrees of freedom in this vector equation, we go into the circular polarisation basis~\cite{geometro},
\begin{align}
    \mathbf{\Psi}^{\pm} = \frac{\Psi_{x} \mp i \Psi_{y}}{\sqrt{2}},\quad\quad |\pm\rangle = \frac{\hat{x} \pm i \hat{y}}{\sqrt{2}},
\end{align}
which allows us to re-express Eq.~(\ref{vectorwaveeq}),
\begin{align}
i\partial_{z'}\Psi^\pm &=- \frac{1}{2\beta}(\nabla'_\perp + i\mathbf{A})^2\Psi^\pm -\frac{\alpha^2\beta r'^2}{2}\Psi^\pm - k(\Delta n')\Psi^\pm \mp \tau\Psi^\pm\\
    &=- \frac{1}{2\beta}(\nabla'_\perp + i\mathbf{A})^2\Psi^\pm -\frac{\alpha^2\beta r'^2}{2}\Psi^\pm - (\Delta\beta \pm\tau)\Psi^\pm\label{eq:circularVWE}
\end{align}
where the equation for the right-circularly polarised eigenvector ${\Psi}^{+}$ corresponds to the top sign, and the equation for ${\Psi}^{-}$ corresponds to the bottom sign. In our fibre, the twist rate $\alpha$ is much greater than the radial position of each core $r$, so the spatial variation of $\tau$($ = \frac{\alpha}{1+\alpha^2 r^2}$) can be neglected, and Eq.~(\ref{eq:circularVWE}) reveals the presence of  circular birefringence. Each of the decoupled equations for each component of the polarisation in this basis is subject to the same vector-potential and scalar-potential terms. The effect of the twist on the polarisation can be described by a relative constant shift of the respective eigenvalues (see Fig.~\ref{fig:s1}\textbf{b} for numerical verification of this effect for the specific geometry of our fibre). By considering only one of the polarisation components, we describe the system using a scalar wave equation of the form
\begin{align}
    i\partial_{z'}\psi' &=- \frac{1}{2\beta}(\nabla'_\perp + i\mathbf{A})^2\psi' -\frac{\alpha^2\beta r'^2}{2}\psi' - k(\Delta n')\psi',\label{eq:scalarwaveeqSI1}\\
    \mathbf{A} &= \alpha\beta(y',-x'),
\end{align}
where the circular birefringence terms have been absorbed as a constant in the index $\Delta n'$. Dropping the $\prime$ symbols for convenience, we arrive at Eq.~(4) in the manuscript,
\begin{align}
    i\partial_{z}\psi &=- \frac{1}{2\beta}(\nabla_\perp + i\mathbf{A})^2\psi -\frac{\alpha^2\beta r^2}{2}\psi - \Delta n(x,y) k \psi
    \label{eq:scalarwaveeqSI}
\end{align}

\subsection{Time-reversal symmetry breaking}
Here, we consider the symmetry properties of the scalar wave equation~(\ref{eq:scalarwaveeqSI1}), and define a natural time-reversal symmetry which is broken due to the twist in our fibre.
The coordinate transform from the laboratory frame to a frame that co-rotates with the waveguides is given by Eq.~(\ref{eq:VWEtransformations2}). In our fibre, the propagation direction, $z$, plays the role of time in the analogous Schr\"{o}dinger equation. We use this analogy to define a time-reversal operator~\cite{Sakurai_1994}:
\begin{equation}
    \hat{\mathcal{T}}:\quad z \longmapsto -z,\,\,\, i \mapsto -i.
\end{equation}

We note that performing the time-reversal operation $\hat{\mathcal{T}}$ on the helicoidal coordinate system,  Eq.~(\ref{eq:VWEtransformations2}), is the same as reversing the twist, $\alpha \mapsto -\alpha$:

\begin{align}
    x'(\alpha) &\longmapsto x\cos{(\alpha [-z])} - y\sin{(\alpha [-z])} =  x\cos{([-\alpha] z)} - y\sin{([-\alpha] z)} \equiv x'(-\alpha)\\
    y'(\alpha) &\longmapsto x\sin{(\alpha [-z])} + y\cos{(\alpha [-z])} =  x\sin{([-\alpha] z)} + y\cos{([-\alpha] z)} \equiv y'(-\alpha)\\
    z' &\longmapsto -z'.
\end{align}

Due to the local $z'$-translation invariance of the Hamiltonian in Eq~(\ref{eq:scalarwaveeqSI1}), the solution $\psi'$ is a function only of the transverse helicoidal coordinates, $x'$ and $y'$. Time reversal then implies
\begin{align}
\psi'(x'(\alpha),y'(\alpha)) \longmapsto \psi'(x'(-\alpha),y'(-\alpha)). \label{eq:zreversalpsi}
\end{align}

When time-reversal is applied to Eq.~(\ref{eq:scalarwaveeqSI1}), $z$-reversal is equivalent to Eq.~(\ref{eq:zreversalpsi}), whereas complex conjugation affects the vector-potential term: 
\begin{equation}
    (\nabla'_{\perp} + i\mathbf{A}(\alpha))^2 \longmapsto (\nabla'_{\perp} - i\mathbf{A}(\alpha))^2 \equiv (\nabla'_{\perp} + i\mathbf{A}(-\alpha))^2  .
\end{equation}
In combination, these transformations imply that Eq.~(\ref{eq:scalarwaveeqSI1}) obeys time-reversal symmetry only if the twist is also reversed simultaneously,
\begin{equation}
    \hat{\mathcal{T}}\big\{\hat{H}(\alpha)\psi'(\alpha)\big\} \longmapsto \hat{H}(-\alpha)\psi'(-\alpha).
\end{equation}\label{Eq:Tsymmetry}
In other words, without the vector potential, the system would be time-reversal symmetric, but the vector potential $\mathbf{A}$ explicitly breaks this time-reversal symmetry.

\subsection{Multicore Twisted Fibre}
Here we derive Eq.~(5) in the main text. When considering a fibre cross-section containing multiple cores, we can use coupled mode theory (equivalently, the tight-binding model) to calculate the fibre's supported supermodes~\cite{22137} using the following equation: 
\begin{align}\label{eq:supermode_nodelta}
\beta \mathbf{u} = \mathrm{C} \mathbf{u},
\end{align}
where $\beta$ is the propagation constant of the supermode, $\mathbf{u}$ is a vector containing $u_{i}$, which is the amplitude of the transverse field profile in the $i$-th core. $\mathrm{C}$ is the coupling matrix that features the individual core $\beta_{0}$ on the diagonal and the coupling coefficients between cores on the off-diagonal. 

In the untwisted case, we can express Eq.~(\ref{eq:supermode_nodelta}) as
\begin{align}\label{eq:supermode}
\Delta \beta \mathbf{u} = \mathrm{C} \mathbf{u},
\end{align}
where $\Delta \beta$ describes the change in propagation constant of a supermode from the propagation constant of an equivalent uncoupled core. By making the change from $\beta$ to $\Delta \beta$, the diagonal of the coupling matrix becomes zero.

To introduce the effects of twist into our supermode equation, we add a twist-dependent change in propagation constant for each core (due to the scalar potential) and we introduce complex phases to the coupling coefficients (due to the vector potential). The change in propagation constant is a consequence of geometry -- as a fibre is twisted, light on the outside travels further to cover the length of the fibre. This change in path length gives rise to a change in propagation constant that is dependent on each core's radial position~\cite{Russell_2017}. As the change in radial path length is different for each core, we introduce the difference between the twisted and untwisted propagation constant along the diagonal of our coupling matrix.

To reintroduce the vector potential, we use a Peierls phase to modify the coupling of light between cores~\cite{Peierls_1933,Luttinger1951,Fang_Yu_Fan_2012}. The Peierls phase is introduced into each coupling coefficient as, 
\begin{equation}
    C_{mj} \longmapsto e^{i\mathbf{A}\cdot \mathbf{r}_{mj}} C_{mj},
\end{equation}
which mirrors how the effects of a magnetic field are introduced into a tight-binding model of, e.g., graphene~\cite{Lado_2015}. 

We can now express each of the coupled equations as: 
\begin{align}\label{eq:supermode_m}
\Delta \beta_{m} u_{m} = \sum_{j\neq m} e^{i\mathbf{A}\cdot \mathbf{r}_{mj}} C_{mj} u_{j} + D_{m} u_{m},
\end{align}
where $C_{mj}$ is the untwisted coupling coefficient which can be calculated from the overlap between two untwisted cores and $D_{m}$ is the change in propagation constant for a given core when it is twisted. 

\section{Real-Space Chern Marker Calculation}
In this section, we define the real-space Chern marker, compute the Chern marker as a function of twist amplitude, and show that in our fibre, there are two band gaps characterised by opposite Chern numbers, in analogy with graphene in a magnetic field.

As highlighted in the main text, Chern numbers are typically defined in systems that feature the translational symmetry of a periodic lattice. Due to this symmetry, a repeating unit cell and reciprocal space can be defined. In such systems, the Chern number is the degree of the map between the reciprocal space torus and the unit sphere on which the eigenvectors of the Hamiltonian (or, in the photonic case, the coupling matrix) lives. Although usually defined in periodic systems, it is also well established that systems without translational invariance can possess non-zero topological invariants, such as the Chern numbers, which can be computed using real-space methods~\cite{KITAEV20062,Mitchell2018}. 

Calculating the Chern number of a system without translational symmetry requires a method that works in real space, such as the Kitaev sum. The Kitaev sum measures a band's integrated chirality within a finite-size system. By defining a projection operator, $\mathcal{P}$, which selects states above a band-gap defined by a cutoff propagation constant, $\beta_{c}$, a non-integer-valued approximation of the local Chern marker $\mathcal{C}$ can be obtained. Following the derivation laid out in Refs.~\cite{KITAEV20062,Mitchell2018}, we arrive at the definition for real-space Chern marker: 
\begin{equation}\label{kitaevapprox}
    \mathcal{C} (\mathcal{P}) = \sum_{j\in A} \sum_{k \in B} \sum_{l \in C} 12 \pi i(\mathcal{P}_{jk}\mathcal{P}_{kl}\mathcal{P}_{lj} - \mathcal{P}_{jl} \mathcal{P}_{lk} \mathcal{P}_{kj}),
\end{equation}
where $\mathcal{P}=\sum_{\beta > \beta_{c}} |u_{\beta}\rangle \langle u_{\beta}|$ is a projection operator that maps modes above a given propagation-constant cutoff to themselves, and modes below this cutoff to zero. In this notation, $\mathcal{P}_{jk}$ is the $(j,k)$th element of the projection operator, where the indices label a single core. The regions $A,B,C$ are three non-overlapping regions defined on the lattice in real space, and are shown on the lattice inset in Fig.~\ref{fig:s7}\textbf{b}. 
\begin{figure}[b]
    \centering
    \includegraphics[width=\linewidth]{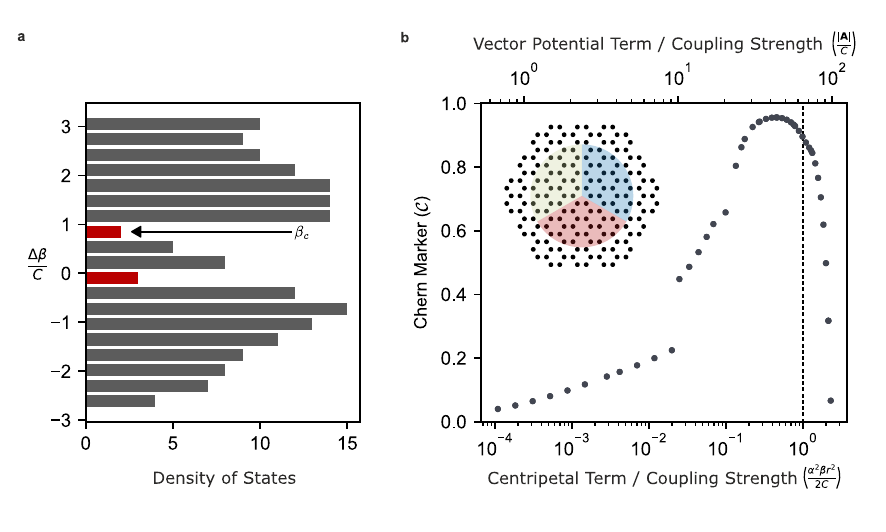}
    \caption{Real-space Chern marker calculation and effect of twist-induced scalar potential. \textbf{a}, The propagation constant density of states is calculated using tight-binding numerics. We label the propagation constant cut-off $\beta_{c}$, which is used to compute the real-space Chern marker shown in \textbf{b}. \textbf{b}, Real-space Chern marker of the upper band (above the cutoff shown in \textbf{a}) is plotted as a function of twist, here non-dimensionalised in two ways: as ratios of the vector potential (top horizontal axis) and the centripetal scalar potential  (bottom horizontal axis) and the  coupling strength $C$ (at zero twist). The graph shows that for both small twist rates and large twist rates, the fibre is topologically trivial, but for intermediate values of the twist rate, the Chern marker indicates non-trivial topological character. The real-space Chern marker in \textbf{b}  and density of states in \textbf{a} have been calculated for the lattice shown in the inset of \textbf{b}.} 
    \label{fig:s7}
\end{figure}

Using this real-space method, we investigate the topology of our multicore fibre system. By measuring the value of the Chern marker for the upper band (above $\beta_{c}$ in Fig.~\ref{fig:s7}\textbf{a}), we gain insight into how the topology changes as a function of twist rate. 
Figure~\ref{fig:s7}\textbf{b} shows how the calculated Chern marker, computed for the lattice shown in the inset, changes as function of twist. 
The graph shows that for both small twist rates and large twist rates, the fibre is topologically trivial, but for intermediate values of the twist rate, the Chern marker indicates non-trivial topological character. For this plot, the nearest neighbour coupling (in the untwisted case) is \SI{4135}{m^{-1}}, which corresponds to the experimentally fabricated fibre at an optical wavelength of \SI{1}{\micro{}m}. The bottom horizontal axis is labelled with the non-dimensionalised amplitude of the centripetal term and the top horizontal axis is labelled with the corresponding non-dimensionalised amplitude of the vector potential, both rescaled by the coupling strength. The twist amplitude simultaneously sets the values along both of the horizontal axes. As the twist increases (i.e., as we move to the right along the horizontal axis) both the centripetal and the vector potential terms increase, and the Chern marker becomes non-zero. However, once the centripetal term becomes greater than the coupling strength (dotted line), the topology breaks down and the Chern marker rapidly decreases. 

So far, we have only considered a single Chern marker computation for the whole of the band structure, by considering only a single band gap (and a single cutoff) for the projection operators. In some models of Chern insulators such as the Haldane model, this is sufficient because there is only a single band gap, with the bands above and below the gap characterised by equal and opposite Chern numbers. However, our fibre is more analogous to the Landau levels in graphene, and we instead observe not only an upper and a lower band, but also a small middle band of states, which creates two separate band gaps. To calculate the Chern numbers of each band independently, we scan the cutoff $\beta_{c}$ across all values of the propagation constant. 

Figure~\ref{fig:s8} shows the Chern numbers for each band in a model fibre system that features a complete lattice (no cut out region) and coupling strength (\SI{6182}{m^{-1}}).
Figure \ref{fig:s8}\textbf{a} shows the calculated Chern number on the horizontal axis, as a function of the cutoff propagation constant $\beta_{c}$ (rescaled by nearest neighbor coupling $C$). As a check, the results in Fig.~\ref{fig:s8}\textbf{a} show that when the cutoff is above or below all of the bands, the Chern marker is zero. If the cutoff is in one of the band gaps, the Chern marker obtains a non-zero value. We observe that inside the lower band gap, the Chern marker is computed to be $-1$, and the Chern marker keeps this near-integer value across the whole plateau corresponding to the band gap.
This is because the edge modes that exist within the band gap are localised to the edge of the lattice, at sites that are not included within the (bulk) Kitaev sum. Crossing the small band of states near $\Delta \beta = 0$, we see that the Chern number flips to $1$, where it remains across the plateau corresponding to the upper band gap. This plot shows that our twisted fibre is characterised by two band gaps with equal and opposite Chern numbers, analogously to graphene in a magnetic field~\cite{Lado_2015}. 

\begin{figure}[tp]
    \centering
    \includegraphics[width=\linewidth]{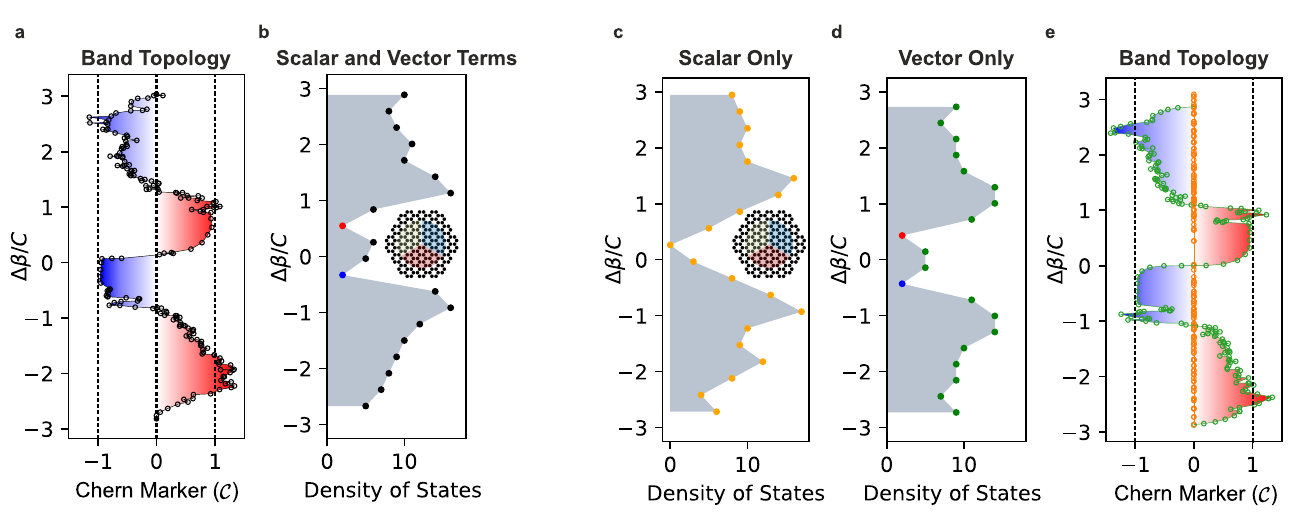}
    \caption{Calculating the real-space Chern marker (which approximates the Chern number) for each band gap present in the band structure of supported propagation constants. \textbf{a}, The Chern marker (for the fibre cross-section shown in the inset of \textbf{b}) is plotted on the $x$-axis, with the associated propagation constant cutoff (used to calculate the Chern marker) on the $y$-axis.  \textbf{b}, The density of states of the fibre's supported propagation constants, which shows the location of the two band gaps. The two band gaps have been coloured to represent their respective Chern markers. Red corresponds to a Chern marker of $1$ and blue corresponds to a Chern marker of $-1$. Both centripetal terms and vector potential terms are included in the calculation.  \textbf{c}, Density of states for a twisted system when ignoring the effects of the vector potential. This system does not break time-reversal symmetry and exhibits only trivial Chern markers ($\mathcal{C} = 0$) for all propagation constant cut-off values, which are shown with orange symbols in \textbf{e}. \textbf{d}, Density of states for a twisted system with no centripetal terms, which does exhibit non-trivial Chern marker values. The calculated Chern markers ($\mathcal{C} \ne 0$) are shown in green in \textbf{e}. This figure uses fibre parameters corresponding to the twist rate of \SI{866}{rad \per m}, and a coupling strength $C$ of \SI{6182}{m^{-1}}. }
    \label{fig:s8}
\end{figure}
\section{Robustness to Disorder}
\begin{figure}[tb]
    \centering
    \includegraphics[width=\linewidth]{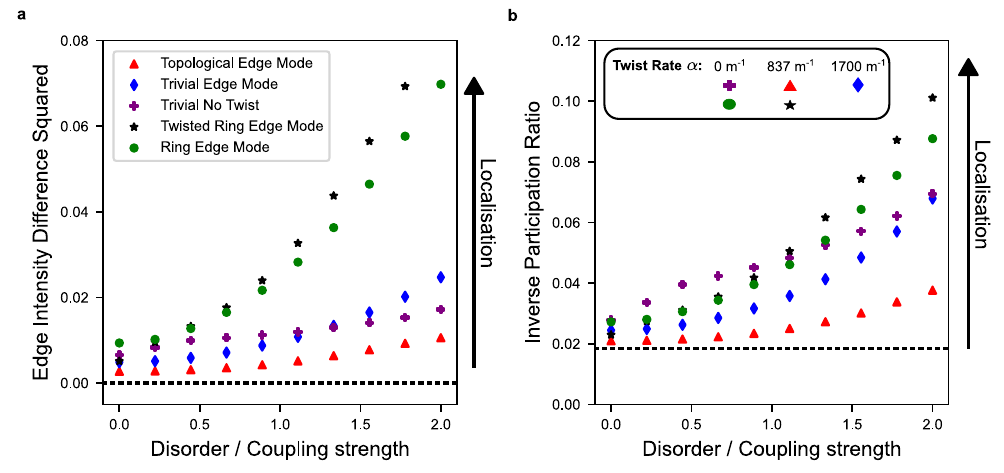}
    \caption{Disorder robustness of different fibre cross-sections. We quantify the disorder localisation that occurs in a fibre cross-section when cores change size or shape and plot this for five different fibre cross-section geometries. \textbf{a}, The Edge Core Intensity Difference is calculated by first finding the intensity difference between a ideal edge mode (equal intensity in all edge cores) and the edge cores of each mode, then squaring and summing these differences to compute a single value. For each fibre cross-section, we compare the mode with the lowest edge core intensity difference in the presence of no disorder. \textbf{b} Same simulation data analysed by computing the Inverse Participation Ratio as a measure of disorder, which leads to the same conclusions as in \textbf{a}. We see that compared to all other fibre structures, the topological edge mode remains the most delocalised (and therefore, closest to an ideal edge mode) across all disorder strengths. }
    \label{fig:s10}
    \end{figure}
    
In this section, we present the details that demonstrate the fibre's topological robustness, as described in the main text. 

We first explore how disorder changes the localisation of the supported supermodes. To compare the localising effects of fabrication-relevant disorder, we augment the coupling matrix of five comparable fibre models (compare to Fig.~4\textbf{b} in the main text, where we include only three of these models): our fabricated topological fibre, an untwisted trivial version of our fabricated fibre, an overly-twisted (\SI{1700}{m^{-1}}) trivial version of our fabricated fibre, an untwisted fibre that only contains cores at the outer edge, and a twisted (\SI{837}{m^{-1}}) fibre only featuring edge cores. In order to make a fair comparison, we ensure the edge cores in all coupling matrices are subject to the same disorder. The disorder is introduced by drawing random values from a uniform distribution between $\pm1/2$, multiplying these values by the disorder strength, and adding this disorder on the diagonal of the coupling matrix in our coupled mode theory (i.e., tight-binding model). Diagonalising the new coupling matrix reveals the system's supermodes, and from these modes, we calculate the localisation plotted in the main text. In the main text, the three systems where we compute this localisation are: (1) a topological edge mode in the fabricated fibre model, (2) an initially delocalised mode in the ring fibre, and (3) a trivially localised edge mode in the \SI{1700}{m^{-1}} twist fibre.

To quantitatively assess the effect of disorder on the supported supermodes, we calculate an intensity difference between an ideal edge mode (where the intensity is uniformly distributed across all cores) and the supermodes supported by each fibre model. We take the difference in intensity at each edge core, square the values so that positive and negative changes do not cancel out, and sum over all of these residuals corresponding to the squared intensity difference. In the main text, we dub this value the Edge Core Intensity Difference, and the modes with the lowest value are the closest to an ideal edge mode. We first use this metric to find the most uniform edge modes in each system for the case with no disorder (corresponding to the most ideal edge state) and plot how these modes' Edge Core Intensity Difference changes as a function of disorder. To ensure we are seeing general effects, we consider $1000$ realisations of $10$ disorder strengths.

During fibre fabrication, fluctuations in parameters (such as temperature) during the drawing process can introduce disorder into any core in the cross-section. Due to the robust nature of topological edge modes, their supported propagation constants change less than trivial bulk modes when exposed to disorder in the lattice. To explore this, we start with the density of states of our model fibre, and introduce random disorder along the diagonal of the coupling matrix. We draw random values from a uniform distribution between $\pm1/2$. We then multiply these values by the disorder strength and add these values to the diagonal of the coupling matrix. Once the new coupling matrix is defined, we calculate the density of states. This is repeated and averaged over $2000$ iterations to show  both the average change in the density of states and their standard deviations. As we report in the main text, for small disorder, the standard deviation in the density of states of the topological edge modes (red and blue) is much smaller than the deviations present in the bulk modes (black). As the magnitude of disorder is increased up to the coupling strength $C$, we see that the topological protection of the edge modes begins to break down and the standard deviation of the density of states becomes comparable to that of the bulk modes.

\end{widetext}

\end{document}